\documentclass[aps,prd,onecolumn,nofootinbib,superscriptaddress]{revtex4}
\pdfoutput=1
\usepackage{graphicx}
\usepackage{amsmath}
\usepackage{amsfonts}
\usepackage{amssymb,ulem}
\usepackage{color}%
\usepackage{dcolumn}

\usepackage{MnSymbol,wasysym}
\usepackage{braket}
\usepackage{eurosym}
\usepackage{calrsfs}
\usepackage[usenames,dvipsnames,svgnames]{xcolor}

\newcommand{\RNum}[1]{\uppercase\expandafter{\romannumeral #1\relax}}
\usepackage[colorlinks=true,linkcolor=blue,urlcolor=blue,filecolor=black,citecolor=red,pdfstartview=FitV,pdftitle={},pdfsubject={},pdfkeywords={},pdfpagemode=None,bookmarksopen=true]{hyperref}
\usepackage[title]{appendix}

\begin{document}
\baselineskip=0.5 cm

\title{Shadow revisiting and weak gravitational lensing with Chern-Simons modification}
\author{Yuan Meng}
\email{mengyuanphy@163.com}
\author{Xiao-Mei Kuang}
\email{xmeikuang@yzu.edu.cn (corresponding author)}
\author{Xi-Jing Wang}
\email{xijingwang@yzu.edu.cn}
\author{Jian-Pin Wu}
\email{jianpinwu@yzu.edu.cn}
\affiliation{Center for Gravitation and Cosmology, College of Physical Science and Technology, Yangzhou University, Yangzhou, 225009, China}

\date{\today}

\begin{abstract}
\baselineskip=0.5 cm
Dynamical Chern-Simons (dCS) gravity has been attracting plenty of attentions due to the fact that it is a parity-violating modified theory of gravity that corresponds to a well-posed effective field theory in weak coupling approximation. In particular, a rotating black hole in dCS gravity is in contrast to the general relativistic counterparts.
In this paper, we revisit the shadow of analytical rotating black hole spacetime in dCS modified gravity, based on which we study the shadow observables, discuss the constraint on the model parameters from the Event Horizon Telescope (EHT) observations, and analyze the real part of quasi-normal modes (QNMs) in the eikonal limit. In addition, we explore the deflection angle in weak gravitational field limit with the use of Gauss-Bonnet theorem. We find that the shadow related physics and the weak gravitational lensing effect are significantly influenced by the CS coupling, which could provide theoretical predictions for a future test of the dCS theory with EHT observations.
\end{abstract}

\maketitle

\section{Introduction}

Einstein's general relativity (GR) is a successful theory in modern physics and passes lots of tests in astrophysics as well as astronomy, however, it is still facing some challenges, such as the explanation of the Universe expansion history, the large scale structure and the understanding of the quantum gravity and so on. Thus, physicists are very interested in finding alternative or higher dimensional theories of gravity to explain our Universe or to further understand the gravity. In particular, recent achievements on astronomical observations, such as the detection of the gravitational waves released from binary black hole mergers \cite{LIGOScientific:2016aoc} and shadows of supermassive black holes \cite{EventHorizonTelescope:2019dse,EventHorizonTelescope:2022xnr}, make it possible to test alternative theories of gravity.  These developments dredge a powerful venue to distinguish or constrain black holes deviating from those in GR.

Since the EHT observation of the M87* black hole was published, the shadow of black holes,  as an important astronomical observation of strong gravitational lensing,  has received considerable attention. In history, the work of shadow was started by Synge \cite{Synge:1966okc} and Luminet \cite{Luminet:1979nyg}, who first gave the photon capture region of the Schwarzschild black hole. After that, Bardeen considered the shadow of the Kerr black hole, and found that the shadow cast of the rotating Kerr black hole is not a standard perfect circle \cite{bardeen1973houches}. Since the shadow of black hole can effectively reflect the information of strong field regime,  it is {extensively investigated} in GR and modified theories of gravity (MoG), see   \cite{Cunha:2018acu,Perlick:2021aok,Chen:2022scf} for reviews. Later, it was proposed in \cite{Hioki:2009na,Kramer:2004hd} that  the deformation and size of shadow could determine  the black hole parameters, {such that it can be further} used to test different theories of gravity, see for examples \cite{Wei:2013kza,Tsupko:2017rdo,Allahyari:2019jqz,Hou:2021okc,Gan:2021pwu,Khodadi:2020jij,Badia:2021kpk,Meng:2022kjs,
Kuang:2022xjp} and references therein.
Considerable studies indicate that the shadow related observables can diagnose and even constrain black hole parameters in alternative gravity theories, though some constraints are still rough.
On the other hand,  during the black hole mergence, the final ringdown phase describes a perturbed black hole
radiating gravitational wave with quasi-normal mode (QNM), which is one of the characteristic properties of the black hole and could provide near horizon information of the black holes. {The QNM frequencies should typically be extracted from the perturbation theory}. But Cardoso et al. proposed that in the eikonal limit, the real part
of QNM frequency is connected with the angular velocity of the circular null geodesics while the imaginary part is connected with the Lyapunov exponent \cite{ Cardoso:2008bp}, which is also valid in rotating black holes. Then the real part of the QNM in the eikonal limit was further related to the shadow radius of static black hole as $\omega_{Re}=\lim\limits_{\ell\gg 1}\frac{\ell}{R_{sh}}$\cite{Jusufi:2019ltj,Liu:2020ola}, and more recently this  connection was extended into the rotating black holes \cite{Jusufi:2020dhz}. This connection may stem from the fact that the gravitational waves could be treated as photons or other massless particles propagating along the last timelike unstable orbit out to infinity, {but more research is needed to fully comprehend this correspondence.}

One particular modified theory of gravity  is the Chern-Simons (CS) gravity which {is frequently} used to discuss quantum gravity and black holes in an {exclusively} theoretical setup. The CS gravity was characterized by the coupling between a scalar field  $\varphi$ and the first-class Pontryagin density  $\ast R_{\mu \nu \rho \sigma }R^{\mu \nu \rho \sigma }$ with the dual Riemann tensor $\ast R_{\alpha \beta \rho \sigma }=\frac{1}{2}\varepsilon_{\rho \sigma }^{~~~ \mu \nu}R_{\alpha \beta \mu \nu }$  \cite{Alexander:2009tp}. This term is dubbed CS term which is parity-violating.  There exist two types of CS modified gravity. The first is the non-dynamical CS (ndCS) gravity, in which the scalar field is a priori prescribed function so that $\ast R_{\mu \nu \rho \sigma }R^{\mu \nu \rho \sigma }$ vanishes to ensure the diffeomorphism invariance of the theory, but the CS term have non-trivial contribution on the solution \cite{Alexander:2007kv}. The second is dynamical CS (dCS) gravity, in which the scalar field is treated as a dynamical field such that the CS term evolves in terms of the field equations \cite{Yunes:2009hc}. It should be emphasized that despite the fact that the ndCS gravity action can be produced from a specific limit of the dCS action, they are typically viewed as two nonequivalent theories.

The dCS gravity theory  is  widely considered in a phenomenological context in cosmology and relativistic
astrophysics, so we shall concentrate on this theory with the action
\begin{equation}
S=\kappa \int d^{4}x\sqrt{-g}R-\frac{1 }{2}\int d^{4}x\sqrt{-g}\left( \nabla \varphi \right) ^{2}+\frac{\gamma }{4}\int d^{4}x\sqrt{-g} \varphi \, \ast R_{\mu \nu \rho \sigma }R^{\mu \nu \rho \sigma }, \label{eq:action}
\end{equation}
where $\kappa=(16\pi G)^{-1}$  and $\gamma$ is the CS coupling constant.
Regarding to the black hole solution in dCS theory, the Schwarzschild black hole of GR is also a solution to the dCS theory  because this theory is parity-violating in the sense that modification from GR due to the
dCS term only appears in  systems with {broken}  parity symmetry \cite{Cardoso:2009pk}, but the rotating black hole would be modified and the rotating black hole with scalar hair emerges.  The analytical spinning black hole  solution with first order in small rotation  was proposed  in \cite{Yunes:2009hc,Konno:2009kg} , which has  been extended to arbitrary order in small rotation \cite{Cano:2019ore}.  A non-perturbative numerical solution for spinning black hole in dCS gravity was constructed by directly solving the {fields' equations of motions }\cite{Delsate:2018ome}.

The leading order correction to the slowly rotating solution  reads as \cite{Yunes:2009hc,Konno:2009kg}
\begin{eqnarray}\label{eq-slowly-metric}
ds^2&&=g_{tt}(r,\theta)dt^2+2g_{t\phi}(r,\theta)dtd\phi+g_{rr}(r,\theta)dr^2+g_{\theta\theta}(r,\theta)d\theta^2+g_{\phi\phi}(r,\theta)d\phi^2.\\
&&=ds^2_{SK}+\frac{5\gamma^2}{4\kappa r^4}(1+\frac{12M}{7r}+\frac{27M^2}{10r^2})a\sin^2\theta dtd\phi,\label{eq-metric}
\end{eqnarray}
where $ds^2_{SK}$ is the metric for slowly rotating Kerr black hole,
\begin{eqnarray}
ds^2_{SK}=&-\left(B+\frac{2a^2 M}{r^3}\cos^2\theta\right)dt^2+\frac{1}{B^2}\left(B-\frac{a^2}{r^2}(1-B \cos^2\theta)\right)dr^2+(r^2+a^2 \cos^2\theta)d\theta^2
\nonumber\\&-\frac{4M}{r}a\sin^2\theta dtd\phi+\left(r^2+a^2(1+\frac{2M}{r}\sin^2\theta)\right)\sin^2\theta d\phi^2, \label{eq-metric-Kerr}
\end{eqnarray}
with  $B=1-2M/r$.  Meanwhile, the  scalar field $\varphi$  has the configuration $\varphi=\left(\frac{5}{2}+\frac{5M}{r}+\frac{9M^2}{r^2}\right)\frac{\gamma a\cos\theta}{4M r^2}$.
It is obvious that as  $\gamma\to 0$, the modified term in \eqref{eq-metric} vanishes and the solution goes back to the slowly rotating Kerr black hole.
The superradiant spectrum in dCS gravity in the context of the slowly rotating black hole has been analyzed in \cite{Alexander:2022avt}.
The optical phenomena in strong field regime of this slowly rotating black hole, such as the flux and the emission spectrum of the accretion disks \cite{Harko:2009kj}, the shadow cast \cite{Amarilla:2010zq}, geodetic precession and strong gravitational lensing \cite{Chen:2010yx}, have partly disclosed {how the CS correction affects} the observational {signals}.

The aim of this paper to further study the shadow observables and the gravitational lensing effect of the slowly rotating black hole in dCS gravity. Our complementary research  to  these issues could help to improve our understanding on the effect of this theory and provide theoretical prediction of the test of the dCS theory with future {more precise} EHT observations.
The remaining of this paper is organized as follows.  In section \ref{sec:shadow and QNM},
 by revisiting the shadow cast, we will check the possible constraint on the CS coupling from the EHT observation of the supermassive black holes, and then analyze the
QNM frequencies in  eikonal limit.   In section \ref{sec:weak lensing}, we calculate the gravitational deflection angle in weak field limit with the use of Gauss-Bonnet theorem.  Section \ref{sec:conclusion} contributes to our conclusion and discussion. We shall use the unit $G=c=1$ and denote the dimensionless quantities $a/M\to a$ and $\gamma/M^2\to \gamma$ for simplification unless otherwise noted.

\section{Shadow cast and quasi-normal modes}\label{sec:shadow and QNM}

The achievement of the EHT is remarkable since it opens a new window to explore the strong gravity regime through the direct observation.   It is known that {the core physics in such observation is the light deflection by the gravitational field.}  In the strong gravity regime, there is a photon region of black hole where the light rays from the light source get captured and it  provides key properties in the black hole shadow. The existence of unstable photon regions outside the event horizon provides the possibility to observe the black hole directly. The photons that escape from the spherical orbits form the boundary of the dark silhouette of the black hole. {Then according to the outside communicators, this dark silhouette is known as black hole shadow}. In particular,
the information of the first black hole image of M87*  published by EHT collaborations in 2019 \cite{EventHorizonTelescope:2019dse}  gives constraints on some shadow observables. Furthermore, from the second black hole image of SgrA*, we know that the angular shadow diameter is evaluated as $d_{sh}=48.7 \pm 7 \mu as$ \cite{EventHorizonTelescope:2022xnr}.
The shadow related physics from EHT observations {is} consistent  with the prediction from the Kerr black hole geometry in GR, but they cannot exclude other black holes in GR or some exotic black holes in MoG. Alternatively,  the EHT observations of shadow can be applied as a tool to constrain the black hole parameters in various theories of gravity.

A preliminary image of the shadow cast of slowly rotating  black hole \eqref{eq-metric} in dCS gravity was studied in \cite{Amarilla:2010zq}, in which  how the dCS coupling parameter affect  the shadow size  $R_s$ and distortion $\delta_s$ were also explored.
{Since in general  $R_s$ and $\delta_s$ may not adequately reflect the  black hole shadow. } Here we shall {broaden our analysis to include} the shadow area  and oblateness.
Then we will connect the theoretical study on the  angular shadow diameter by the slowly rotating  black hole with the EHT observations of the supermassive  black holes, and discuss the {potential} constraint on the dCS coupling parameter.

\subsection{Shadow cast}

 To determine the apparent shape of the shadow of the black hole, one usually constructs the celestial coordinates $(\alpha, \beta)$, in which  $\alpha$  is the apparent perpendicular distance of the image as seen from the rotating axis of  symmetry and  $\beta$ is  the apparent perpendicular distance of the image from its projection on the equatorial plane.
For the distant observer,  the shadow boundary is described in celestial coordinates as \cite{Abdujabbarov:2016hnw}
\begin{eqnarray}
\alpha=\lim_{r_o\to\infty}\left(-r_o^2\sin\theta_o\frac{d\phi}{dr}\Big|_{(r=r_o,\theta=\theta_o)}\right), \quad  \beta=\pm\lim_{r_o\to\infty}\left(r_o^2\frac{d\theta}{dr}\Big|_{(r=r_o,\theta=\theta_o)}\right).
\end{eqnarray}
Here $(r_o,\theta_o)$ {denotes} the observer's position in Boyer-Lindquist coordinate. $r_o$ {measures} the far distance from the observer to  black hole while $\theta_o$ is the inclination angle between the line of sight of the observer and the spinning axis of the black hole.

To study the shadow cast of the slowly rotating  black hole with CS correction \eqref{eq-metric}, we first derive the null geodesic equations and  two  constants of motion, namely $\xi$ and $\eta$. The details  are present in Appendix \ref{sec:app-1}. Then, we further calculate  $d\phi/dr$ and $d\theta/dr$, and reduce the celestial coordinates as
\begin{eqnarray}\label{eq:X-Y}
\alpha(r_p)=-\xi(r_p)\csc\theta_o,~~~
\beta(r_p)=\pm\frac{a^{2}+4(\xi(r_p)^{2}+\eta(r_p))+a^{2}\cos2\theta_{0}-4\xi(r_p)^{2}\csc^{2}
\theta_{0}} {4\sqrt{\eta(r_p)-\xi(r_p)^{2}\cot^{2}\theta_{0}}},
\end{eqnarray}
where  $\xi(r_p)$ and $\eta(r_p)$ are also known as the impact parameters {depending on} the radius $r_p$ of circular photon orbits in the photon region. It is obvious that  $\xi(r_p)$ and $\eta(r_p)$  determine the shape of black hole shadow.

\begin{figure} [h]
{\centering
\includegraphics[width=2in]{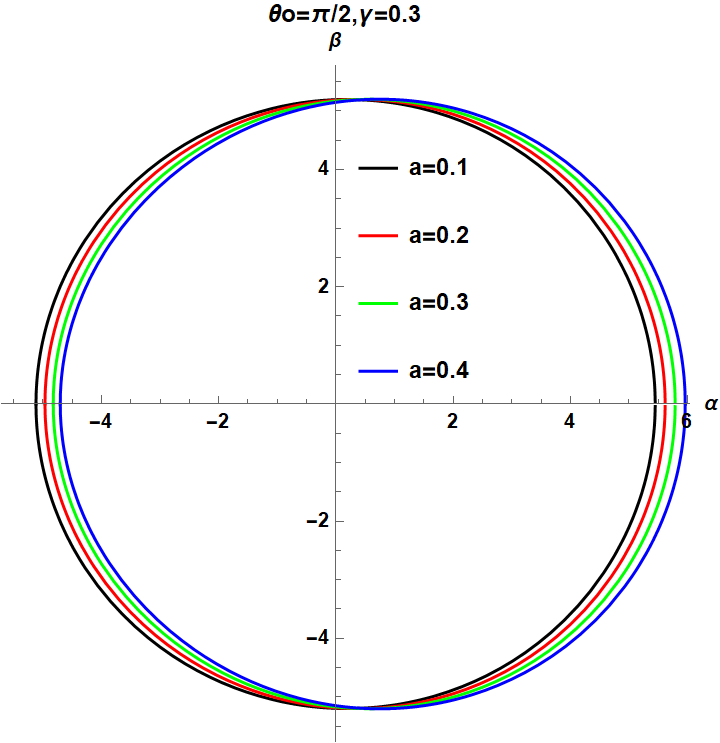}\hspace{0.5cm}
\includegraphics[width=2in]{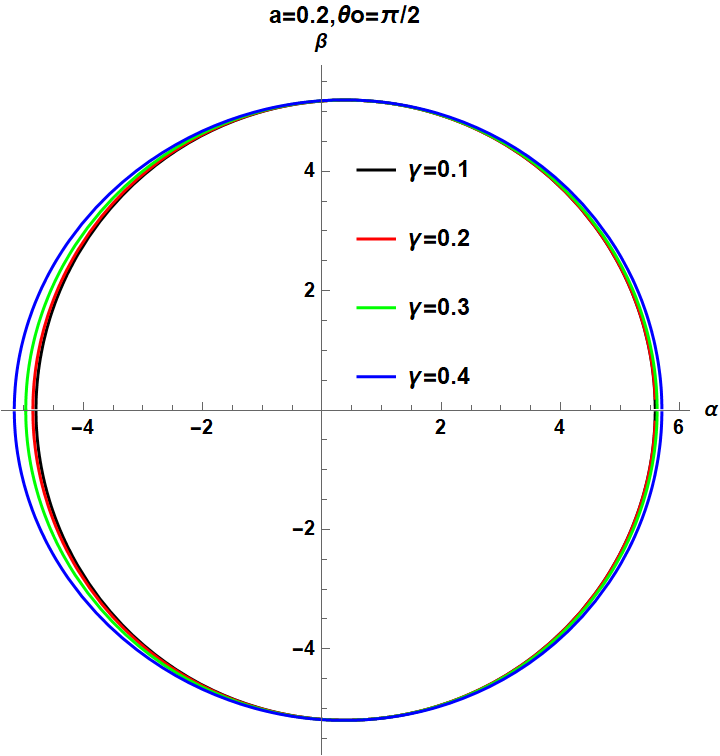}
\includegraphics[width=2in]{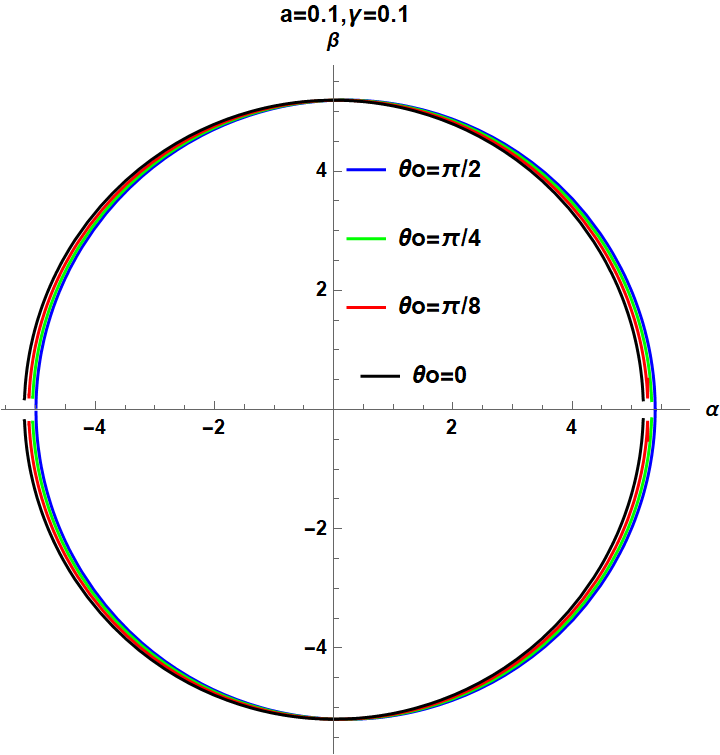}
   \caption{Black hole shadow seen by an observer at infinity distance for various parameters. Here, we have set $M=1$ in the calculations.}   \label{fig:shadow}}
\end{figure}

The shadow boundary for the observer at spatial infinity is depicted  in Fig. \ref{fig:shadow}.
It is noticed that for the slowly spinning black hole in dCS gravity, the spin parameter, $a$, and the CS coupling parameter, $\gamma$, should be far less than the black hole mass, so the rescaled $a$ and $\gamma$ are far smaller than $1$. However, here in order to better reflect the influence of  $a$ and  $\gamma$ on the black hole shadow, we have taken $a=\gamma=0.4$ as the maximum in the calculation. In the left and right panels, we can see that with the increase of parameter $a$ and the inclination angle $\theta_o$, the shadow  will move to the right while its deformation is negligible.  In the middle panel, it is clear that with the increase of the coupling parameter, the shadow becomes large.
\begin{figure} [h]
{\centering
\includegraphics[width=2in]{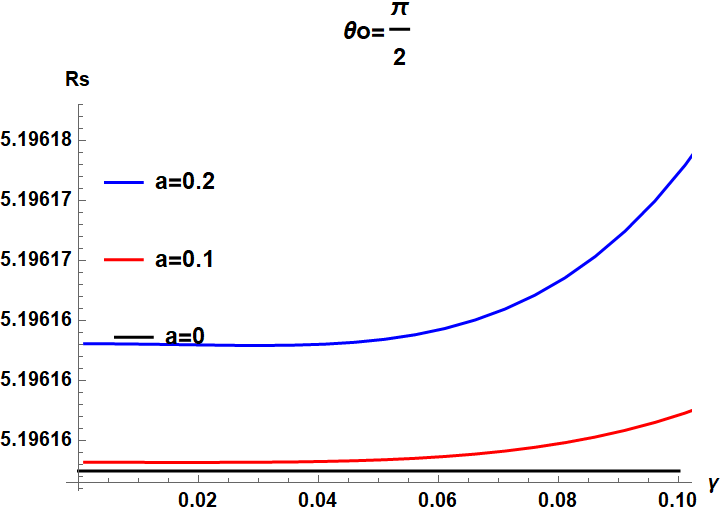}\hspace{1cm}
\includegraphics[width=2in]{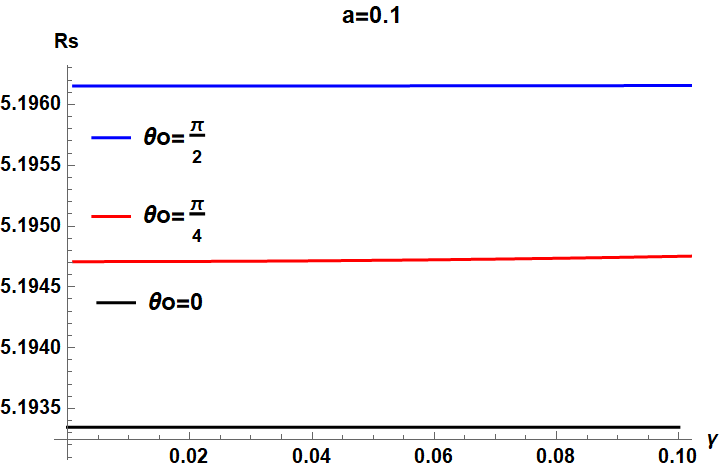}\hspace{1cm}
   \caption{The radius of the reference circle for the shadow, $R_s$, as a function of CS coupling parameter. Here we have set $M=1$.}   \label{fig:Rs-a-g}}
\end{figure}
\begin{figure} [h]
{\centering
\includegraphics[width=2in]{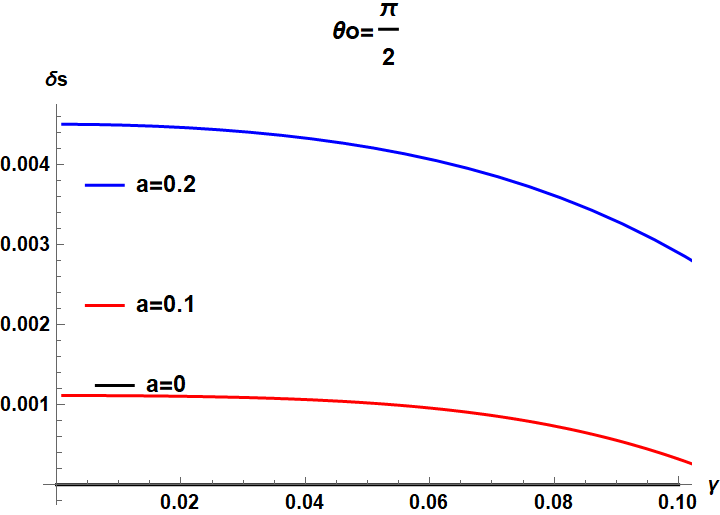}\hspace{1cm}
\includegraphics[width=2in]{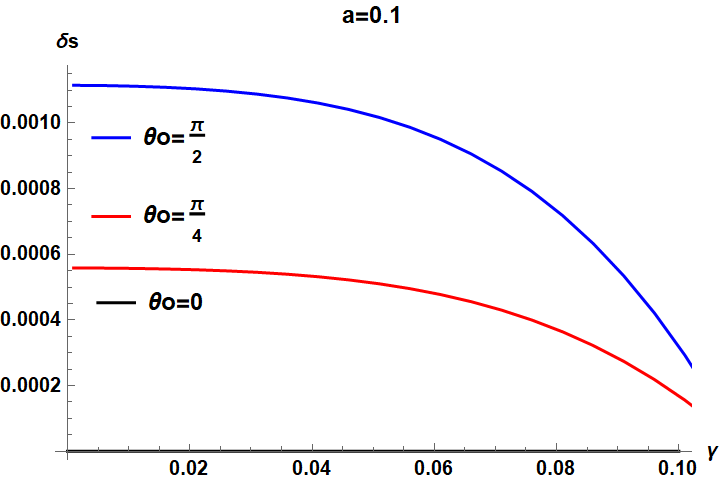}\hspace{1cm}
   \caption{The distortion for the shadow, $\delta_s$,  as a function of CS coupling parameter. Here we have set $M=1$.}   \label{fig:deltaS-a-g}}
\end{figure}

\begin{figure} [h]
{\centering
\includegraphics[width=2in]{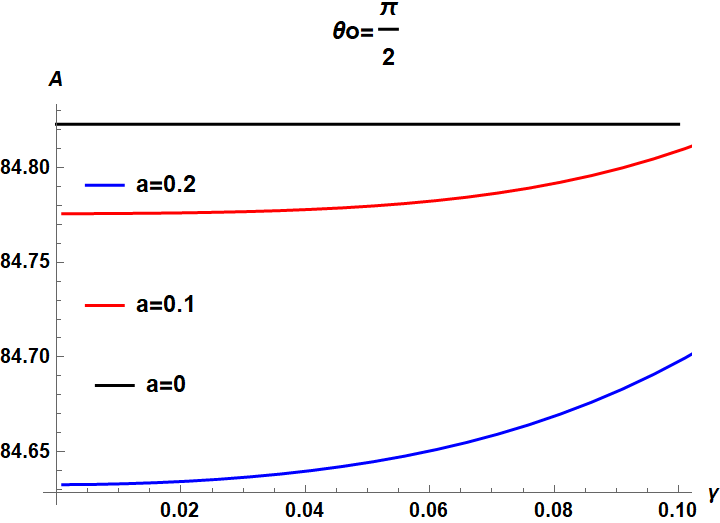}\hspace{1cm}
\includegraphics[width=2in]{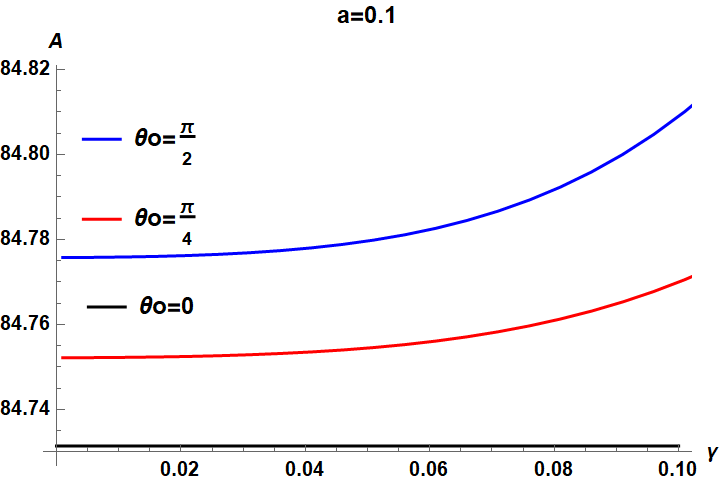}\hspace{1cm}
   \caption{The area for the shadow, $A$, as a function of CS coupling parameter. Here we have set $M=1$.}   \label{fig:A-a-g}}
\end{figure}

\begin{figure} [h]
{\centering
\includegraphics[width=2in]{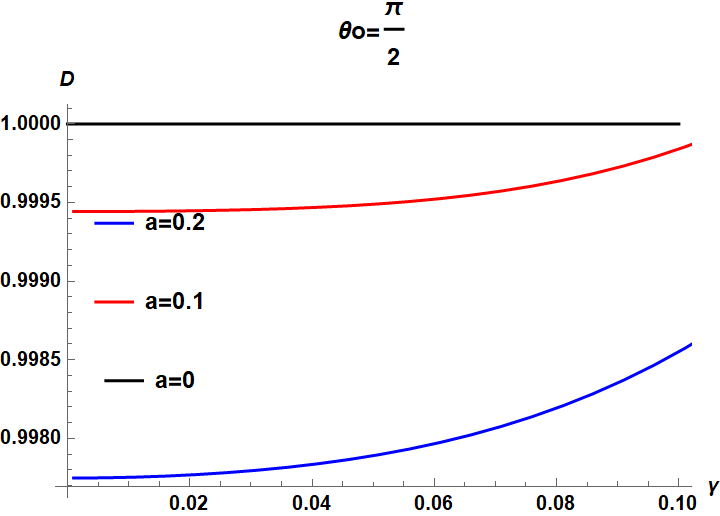}\hspace{1cm}
\includegraphics[width=2in]{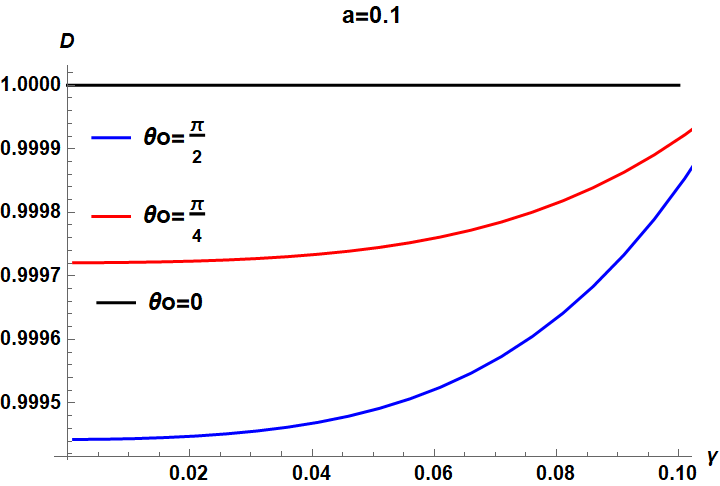}\hspace{1cm}
   \caption{The oblateness for the shadow, $D$, as a function of CS coupling parameter. Here we have set $M=1$.}   \label{fig:Ds-a-g}}
\end{figure}

In order to quantitatively describe the properties of shadow size and distortion that we see from Fig. \ref{fig:shadow}, we denote the top, bottom, right and left of the shadow boundary as $(X_t, Y_t)$, $(X_b, Y_b)$, $(X_r,0)$ and $(X_l, 0)$, respectively and $(X_l', 0)$ as the leftmost edge of its reference circle. Then we can analyze the two groups of characterized shadow observables: the radius $R_s$ of the reference circle and the deviation $\delta_s$ of the left edge of the shadow from the reference circle defined as \cite{Hioki:2009na}
\begin{eqnarray}
R_s=\frac{\left(X_t-X_r\right)^2+Y_t^2}{2|X_r-X_t|}~, \quad \delta_s=\frac{|X_l-X_l'|}{R_s}~,
\label{eq-Rs}
\end{eqnarray}
{along with}  the shadow area $A$ and oblateness $D$ defined as \cite{Kumar:2018ple}
\begin{eqnarray}
A=2\int_{r_p{}_{\rm min}}^{r_p{}_{\rm max}}\left(Y\left(r_p\right)\frac{dX\left(r_p\right)}{dr_p}  \right)dr_p~, \quad
   D=\frac{X_r-X_l}{Y_t-Y_b}~.\label{eq-A}
\end{eqnarray}
{Since we are interested in the effect of the CS coupling, so we show the aforementioned} shadow observables as {functions} of $\gamma$ in Figs. \ref{fig:Rs-a-g}-\ref{fig:Ds-a-g}. Figs. \ref{fig:Rs-a-g}-\ref{fig:deltaS-a-g} show that the radius $R_s$ (distortion $\delta_s$) increases (decreases) as the CS coupling parameter increases, meaning that the CS scalar field will enlarge the black hole shadow but hold back its deformation.  Both $R_s$ and $\delta_s$ {increase} as we increase the {spinning} parameter $a$ and the inclination angle $\theta_o$ for the slowly spinning black hole.
Meanwhile, Figs. \ref{fig:A-a-g}-\ref{fig:Ds-a-g} show that both the shadow area $A$ and oblateness $D$ increase for larger $\gamma$, implying that the CS scalar field makes the shadow larger and oblater. In addition, larger  spinning parameter $a$ results in black hole shadow with smaller area and oblateness, while larger inclination angle corresponds to shadow with larger area but smaller oblateness. It is noticed that the effect of $\gamma$ should be more prominent for faster spinning cases, but {unfortunately}, the analytical form of a complete black hole solution with general value of $a$ is still missing.

\subsection{Constraint on the parameter from EHT observations}

Then we will
{consider the slowly rotating  black hole with CS correction as M87* and SgrA* black hole, respectively, } and examine the possible constraints on the black hole parameters with the use of EHT observations.
It was shown that the image of supermassive black hole M87* photographed by the EHT is crescent shaped \cite{EventHorizonTelescope:2019dse}, based on which, {the range of spinning parameter of M87* was fixed as} $|a|\geq0.4 $ in the prograde case while  $|a|\geq0.5$ in the retrograde case \cite{Nemmen:2019idv}. This means that the EHT observations will exclude the slowly rotating black hole  as supermassive M87* black hole. So it is not suitable to use the EHT observations on M87* to constrain the model parameters of the slowly rotating solution of gravitational theories.

\begin{figure} [h]
{\centering
\includegraphics[width=2in]{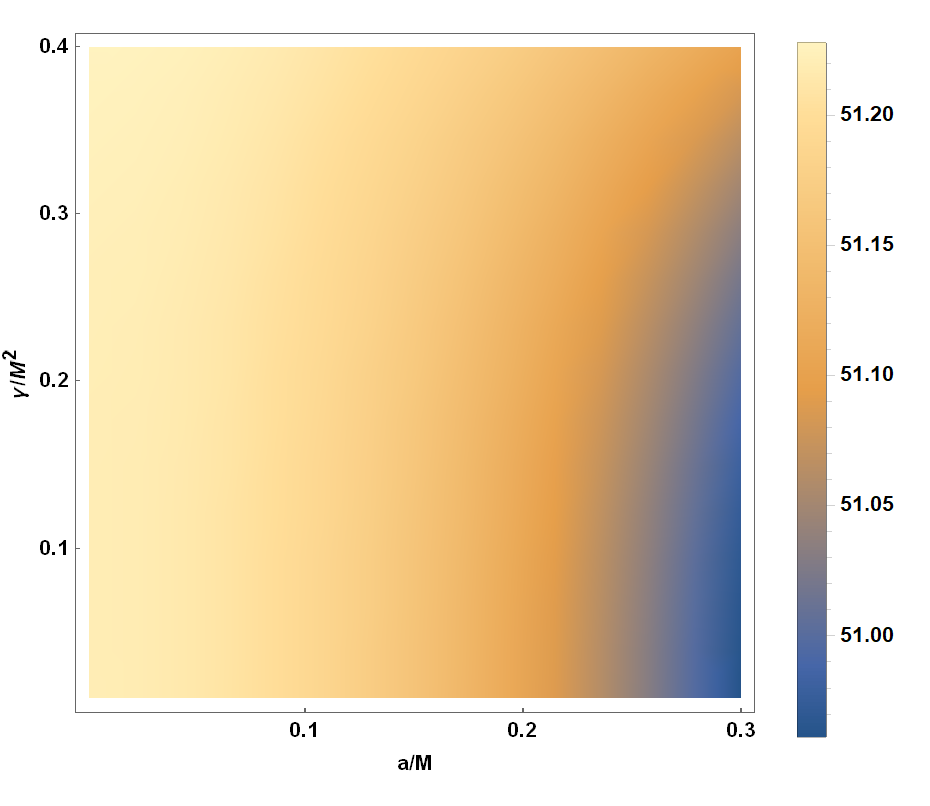}\hspace{1cm}
   \caption{The density plots of the angular diameter $d_{sh}$ by treating the slowly rotating black hole in dCS theory as supermassive Sgr A*. We fix $\theta_o=5^\circ$.}   \label{fig:Sgr-theta-d}}
\end{figure}

On the other hand, the image of supermassive SgrA* black hole  from EHT gives the angular shadow diameter $d_{sh}=48.7 \pm 7 \mu as$ \cite{EventHorizonTelescope:2022xnr}, but its spinning parameter is still unclear. So we can assume the slowly rotating black hole as the candidate {for} supermassive SgrA* black hole { and check} the constraints on the black hole parameters $a$ and $\gamma$ from the EHT observation  $d_{sh}=48.7 \pm 7 \mu as$.
To better refer to the EHT observations, we consider the inclination angle  $\theta_o=\theta_{jet}=5^{\circ}$  as it may be EHT's {favorite}  jet inclination among the options \cite{Issaoun:2019afg}.
The angular diameter $d_{sh}$ of the black hole shadow is given as \cite{Kumar:2020owy}
\begin{equation}
d_{sh}=\frac{2R_A}{d},~~\mathrm{and}~~~R_A=\sqrt{\frac{A}{\pi}},
\label{eq-thetaD}
\end{equation}
where $A$ is the area of the shadow defined in  \eqref{eq-A} and $d$ is the distance {from the SgrA* to the earth.}
Inserting the realistic  $M=4.0\times 10^{6}M_{\odot}$ and $d=8.35 Kpc$ for SgrA*,  we show density plot of the angular diameter in  Fig. \ref{fig:Sgr-theta-d}.
It is obvious that $d_{sh}$ of the slowly spinning black hole in dCS gravity is in good agreement with EHT observation on SgrA*. So in this sense, we can conclude that the EHT observations on  SgrA* black hole shadow cannot rule out the slowly rotating  black hole with CS correction. In addition, the current EHT observations on  SgrA* cannot further constrain the spinning or the dCS coupling parameter. We expect future precise EHT observations can provide more significant information on the constraint of CS term.

\subsection{Shadow radius and QNMs in the eikonal limit}

In dCS gravity, Cardoso et al. found that the CS correction has significant influence on the gravitational perturbation on the Schwarzschild black hole such that the isospectrality was broken \cite{Cardoso:2009pk}; Then the QNMs of gravitational perturbation on the slowly rotating black holes in dCS theory were calculated in analytical method \cite{Srivastava:2021imr} and numerical method \cite{Wagle:2021tam}, respectively, which focused on the small momentum ($\ell\sim 1$) modes. Those studies converge to a conclusion that the QNMs related physics could be used to test dCS theory and even constrain the dCS coupling parameter.
Here we will focus on the QNM frequencies in the eikonal limit ($\ell\gg 1$) and its connection with the shadow radius.  To proceed, we employ the procedure proposed in \cite{Jusufi:2020dhz,Ghasemi-Nodehi:2020oiz} into the slowly rotating metric \eqref{eq-slowly-metric} of dCS gravity in the equatorial plane with $\theta=\pi/2$.
The Lagrangian for photons is written as
\begin{eqnarray}
\mathcal{L}=\frac{1}{2}\left(g_{tt}\dot{t}^2+g_{rr}\dot{r}^2+2g_{t\phi}\dot{t}\dot{\phi}+g_{\phi\phi}\dot{\phi}^2\right),
\end{eqnarray}
from which we can introduce the momenta  as $p_t=-E, ~ p_{\phi}=L_z$ and $p_{r}=g_{rr}\dot{r}$.
Then from the Hamiltonian $H=p_t\dot{t}+p_{\phi}\dot{\phi}+p_r\dot{r}-\mathcal{L}=0$, we can derive the {expression} of $\dot{r}\equiv\mathcal{V}$ which should satisfy the following conditions
\begin{eqnarray}\label{eq-condi}
\mathcal{V}\mid_{r=r_c}=0,~~~~\frac{d \mathcal{V}}{dr}\mid_{r=r_c}=0,
\end{eqnarray}
for the existence of circular geodesic with radius $r_c$. Inserting the metric \eqref{eq-slowly-metric} into above conditions and defining $\mathcal{R}=L_z/E$, we can obtain
\begin{equation}
\mathcal{R}_s^{\pm}=-\left(\frac{g'_{t \phi}\pm\sqrt{(g'_{t\phi})^2-g'_{\phi\phi}g'_{tt}}}{g'_{tt}}\right)\mid_{r=r_0^{\pm}},
\label{eq-Rsa2}
\end{equation}
where $r_0^{\pm}$ {are} the radius of circular null {geodesics} for the retrograde and prograde cases, respectively, determined by {the solutions to the equation}
\begin{equation}
g_{\phi \phi}+2g_{t \phi}\mathcal{R}_s^{\pm}+g_{tt}(\mathcal{R}_s^{\pm})^2=0.
\label{eq-Rsa22}
\end{equation}

\begin{figure} [h]
{\centering
\includegraphics[width=2in]{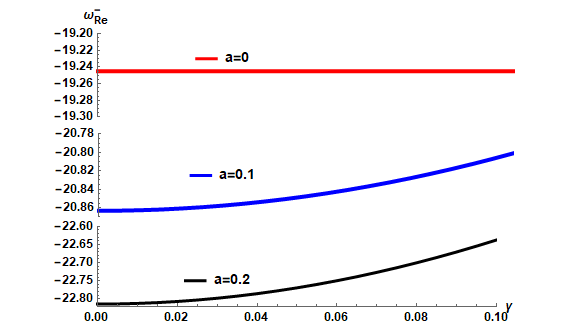}\hspace{1cm}
\includegraphics[width=2in]{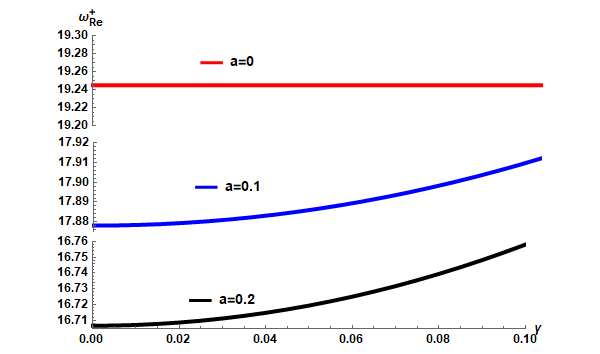}
   \caption{The reals part of QNMs as a function of $\gamma$ for different $a$. Here we fix $M=1$ and $\ell=100$.}   \label{fig:Wa}}
\end{figure}
\begin{figure} [h]
{\centering
\includegraphics[width=2in]{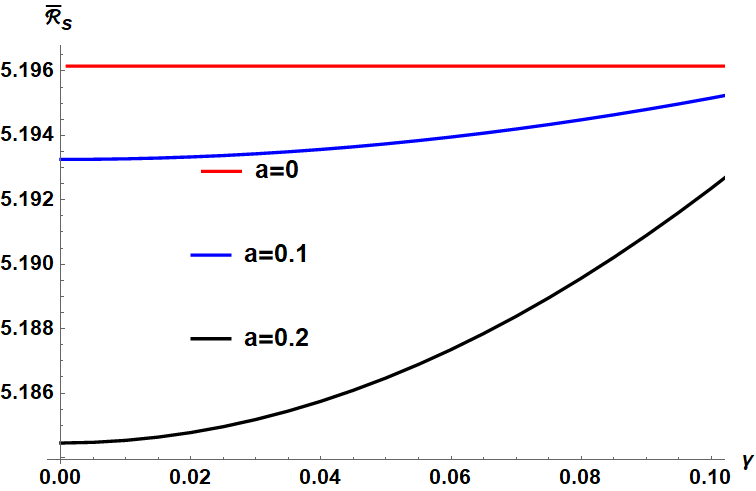}
   \caption{The typical shadow radius for the slowly rotating black hole as a function of $\gamma$. Here we fix $M=1$.}   \label{fig:RSa}}
\end{figure}

Then using the geometric-optics correspondence between the parameters of QNM frequency and the conserved quantities along geodesics in the equatorial plane, we can consider $E\to \omega_{Re}$, $J\to m$ with $m=\pm \ell$ in the eikonal limit where $m$ is  the azimuthal quantum number while $\pm \ell$ is angular momenta of the prograde and retrograde modes, respectively. According to \cite{Jusufi:2019ltj}, we can obtain
\begin{equation}
\omega_{Re}^{\pm}=\lim_{\ell\gg1}\frac{\ell}{\mathcal{R}_s^{\pm}}.
\label{eq-Rsap}
\end{equation}
The real part of QNM frequency, $\omega_{Re}^{\pm}$ with $\ell=100$ is depicted in Fig. \ref{fig:Wa} which shows that both  $\omega_{Re}^{+}$ and $\omega_{Re}^{-}$ {increase} as $\gamma$ increases in the rotating case.
These QNMs frequencies are usually connected with the typical shadow radius for the observer in the equatorial plane, defined as  \cite{Jusufi:2020dhz,Ghasemi-Nodehi:2020oiz}
\begin{eqnarray}
\bar{\mathcal{R}_s}=\frac{\mathcal{R}_s^+-\mathcal{R}_s^-}{2}=\frac{\ell}{2}\left(\frac{1}{\omega_{Re}^{+}}-\frac{1}{\omega_{Re}^{-}}\right).
\end{eqnarray}
We show $\bar{\mathcal{R}_s}$ as a function of $\gamma$ for samples of $a$ in Fig. \ref{fig:RSa}, which indicates that the typical shadow radius increases as the CS coupling parameter {increases} but decreases as the spinning parameter increases. It is noticed that for $a=0$, we see that $\omega_{Re}^{\pm}$ and $\bar{\mathcal{R}_s}$ are {both} independent of $\gamma$ as expected because the CS correction has no {print} on the static black hole solution.

\section{Gravitational deflection in weak filed limit}\label{sec:weak lensing}

It is well known that the light rays will deflect when they propagate in {gravity} field, due to the gravitational lensing effect. This effect mainly depends on the nature of the central source, so it is a useful tool to diagnose theories of gravity or to test exotic object in our Universe. The strong gravitational lensing effect by the slowly rotating black hole in dCS gravity has been well studied in \cite{Chen:2010yx}.
In this section, we extend the study into the weak gravitational lensing and focus on the dCS correction on the gravitational deflection angle of light. To this end, we shall use the Gibbons-Werner method \cite{Gibbons:2008rj}, in which the Gauss-Bonnet theorem in the optical geometry is {resorted}. Specifically,
Gibbons and Werner {proposed} that for a static and spherically symmetric black hole,  the deflection angle of light can be calculated by integrating the Gaussian curvature of the optical metric outwards from the light ray, as a consequence of the focusing of light rays emerges as a global topological effect. Their method was then extended to compute the light deflection {angle} of a static spherical black hole with cosmological constants for the finite distances from the black hole to the light source and the observer \cite{Ishihara:2016vdc}.
Moreover, Gibbons-Werner method was soon used in stationary and axisymmetric black holes \cite{Werner:2012rc}. Calculating the light deflection angle with this method has been widely {investigated} in various spherically symmetric or axisymmetric  black holes.
For a comprehensive review on the applications of the Gauss-Bonnet theorem to gravitational deflection angle of light in weak field limit, one refers to the review work \cite{Ono:2019hkw}, by following the steps of which we will proceed.

\begin{figure} [h]
{\centering
\includegraphics[width=3in]{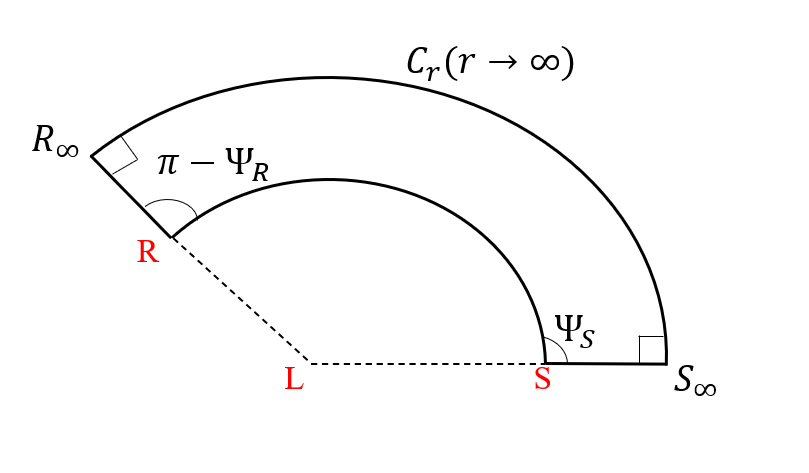}\hspace{1cm}
   \caption{The schematic figure of lensing setup and domain of integration.}   \label{fig:Lensobject}}
\end{figure}

We firstly study the orbit equation of photon on the equatorial plane ($\theta=\pi/2$). For the stationary and axisymmetric spacetime \eqref{eq-slowly-metric}, we mention in the appendix that there exists two constants of motion, $E$ and $L_z$, so we can define the impact parameter $b$ as
\begin{align}
    b\equiv\frac{L_z}{E}=\frac{g_{t\phi}+g_{\phi\phi}\frac{d\phi}{dt}}{-g_{tt}-g_{t\phi}\frac{d\phi}{dt}}.\label{eq:impactEQ}
\end{align}
The null condition $ds^2=0$ leads to the following orbit equation of photon
\begin{equation}
    \left(\frac{dr}{d\phi}\right)^2=\frac{(g_{t\phi}^2-g_{tt}g_{\phi\phi})(g_{\phi\phi}+2g_{t\phi}b+g_{tt}b^2)}{g_{rr}(g_{t\phi}+g_{tt}b)^2},
\end{equation}
where we have inserted the impact parameter \eqref{eq:impactEQ}. For convenience, we introduce $r\equiv1/u$ and then the orbital equation of photon can be rewritten as
\begin{equation}
    \left(\frac{du}{d\phi}\right)^2=\frac{u^4(g_{t\phi}^2-g_{tt}g_{\phi\phi})(g_{\phi\phi}+2g_{t\phi}b+g_{tt}b^2)}{g_{rr}(g_{t\phi}+g_{tt}b)^2}.
    \label{eq-orbital}
\end{equation}
In principle, we can get the orbit trajectory by inserting metric functions \eqref{eq-metric} and \eqref{eq-metric-Kerr} into the \eqref{eq-orbital}. However, due to the complexity, we cannot solve the differential equation analytically. Instead we use the approximations for the weak field, slow rotation and small dCS coupling $\gamma$ contribution to get the analytical approximation solution. The orbital function of the photon is thus obtained as \footnote{Hereafter we will omit the term $\mathcal{O}(M^2, a^2, \gamma^3)$ for simplicity.}
\begin{equation}
    \begin{aligned}
        u\simeq&\frac{\sin\phi}{b}+\frac{M(1+\cos^2\phi)}{b^2}-\frac{2aM}{b^3}+\frac{5a\pi\gamma^2(6(\pi-2\phi)\cos\phi+9\sin\phi+\sin3\phi)}{4b^6}\\
        &+\frac{5aM\pi\gamma^2(2997+1360\cos2\phi-5\cos4\phi-126(\pi-2\phi)\sin2\phi)}{84b^7}+\mathcal{O}(M^2, a^2, \gamma^3).
        \label{eq-orbital1u}
    \end{aligned}
\end{equation}
Meanwhile we solve (\ref{eq-orbital1u}) for $\phi$ obtained as
\begin{align*}
    \begin{split}
        \phi\simeq \left \{
        \begin{array}{ll}
            \arcsin(bu)+\frac{2aM}{b^2\sqrt{1-b^2u^2}}+\frac{M(b^2u^2-2)M}{b\sqrt{1-b^2u^2}}+\frac{5a\pi\gamma^2(-3bu+b^3u^3-3\sqrt{1-b^2u^2}\arccos(bu))}{b^5\sqrt{1-b^2u^2}}\\-\frac{10aM\pi(544-816b^2u^2+204b^4u^4+47b^6u^6)\gamma^2}{21b^6(1-b^2u^2)^{\frac{3}{2}}}                    & (|\phi|<\frac{\pi}{2})\\
            \pi-\arcsin(bu)-\frac{2aM}{b^2\sqrt{1-b^2u^2}}+\frac{M(2-b^2u^2)}{b\sqrt{1-b^2u^2}}+\frac{5a\pi\gamma^2(3bu-b^3u^3+3\sqrt{1-b^2u^2}\arccos(bu))}{b^5\sqrt{1-b^2u^2}}\\
            +\frac{10a\pi M(544-816b^2u^2+204b^4u^4+47b^6u^6)\gamma^2}{21b^6(1-b^2u^2)^{\frac{3}{2}}}                                 & (|\phi|>\frac{\pi}{2})
        \end{array},
        \right.
    \end{split}
\end{align*}
where we can choose the domain of $\phi$ to be $-\pi < \phi \leq \pi$ without loss of generality. As shown in Fig. \ref{fig:Lensobject}, the range of the angular coordinate value $\phi_S$ at the source ($S$) point is $-\pi/2\leq \phi_S < \pi/2$. But for the receiver ($R$) point {, the range of $\phi_R$} is $|\phi_R|>\pi/2$. We assume the infinite distance limit for the source and receiver, namely $u_S, u_R\rightarrow0$, which {leads to} the angles $\phi_S \rightarrow 0$ and $\phi_R \rightarrow \pi$.

For the source and receiver {in} the equatorial plane in the axisymmetric spacetime, the definition of deflection angle is
\begin{equation}
\alpha\equiv\Psi_R-\Psi_S+\phi_{RS}.
\label{eq-ARPH}
\end{equation}
Here $\Psi_S$ and $\Psi_R$ are the included angles of the connecting line between the source and the lens, and the connecting line between the observer and the lens and the radial direction of the light rays respectively. $\phi_{RS}$ is the longitude separation angle between source and observer (cf. Fig.\ref{fig:Lensobject}). In addition, we denote the integral region of the quadrilateral ($R_\infty, R, S, S_\infty$) as $\mathcal{D}$.
According to Gauss-Bonnet theorem, the deflection angle can also be written by
\begin{equation}
\alpha=-\int\int_\mathcal{D}K_o dS-\int^S_R\kappa_gd\ell,
\label{eq-Arpb}
\end{equation}
where $K_o$ and $\kappa_g$ are the Gaussian curvature of the surface $\mathcal{D}$ and the geodesic curvature of light rays, respectively, $dS$ is the area element of surface, and $d\ell$ is infinitesimal line element along the boundary of surface. Now let us compute these two terms {in \eqref{eq-Arpb}}.
To this end, we rewrite $dt$ from the null condition $ds^2=0$ for the metric \eqref{eq-slowly-metric}  as
\begin{equation}
dt=\sqrt{\rho_{ij}dx^idx^j}+\beta_idx^i,
\label{eq-dt3}
\end{equation}
where $i$, $j$ run from 1 to 3, and $\rho_{ij}$ and $\beta_{i}$ are
\begin{equation}
\rho_{ij}dx^idx^j\equiv-\frac{g_{rr}(r,\theta)}{g_{tt}(r,\theta)}dr^2-\frac{g_{\theta\theta}(r,\theta)}{g_{tt}(r,\theta)}d\theta^2+\\
\frac{g_{t\phi}^2(r,\theta)-g_{tt}(r,\theta)g_{\phi\phi}(r,\theta)}{g_{tt}^2(r,\theta)}d\phi^2,
\label{eq-metrspa}
\end{equation}
\begin{equation}
\beta_idx^i\equiv-\frac{g_{t\phi}(r,\theta)}{g_{tt}(r,\theta)}d\phi.
\label{eq-metrspa1}
\end{equation}

Firstly, for the light propagation in the equatorial plane, by considering the 2-dimensional Riemann tensor $^{(2)}R_{r\phi r\phi}$, the Gaussian curvature can be defined as
\begin{equation}
\begin{aligned}
K_o&=\frac{^{(2)}R_{r\phi r\phi}}{\text{det} \rho_{ij}^{(2)}}=\frac{1}{\sqrt{\text{det}\rho_{ij}^{(2)}}}\left[\frac{\partial}{\partial\phi}\left(\frac{\sqrt{\text{det}\rho_{ij}^{(2)}}}{\rho_{rr}^{(2)}}{^{(2)}\Gamma_{rr}^\phi}\right)-
\frac{\partial}{\partial r}\left(\frac{\sqrt{\text{det}\rho_{ij}^{(2)}}}{\rho_{rr}^{(2)}}{^{(2)}\Gamma_{r\phi}^\phi}\right)\right],
\label{eq-Koo}
\end{aligned}
\end{equation}
where $^{(2)}R_{r\phi r\phi}$, $^{(2)}\Gamma_{rr}^\phi$ and $\text{det}\rho_{ij}^{(2)}$ are defined by the optical metric $\rho_{ij}$ on the equatorial plane.
Therefore, the integral of Gaussian curvature over the closed surface is \cite{Ono:2019hkw}
\begin{equation}
-\int\int_\mathcal{D}K_odS=\int^{\phi_R}_{\phi_S}\int^{\infty}_{r_{(\phi)}}K_o\sqrt{\text{det}\rho^{(2)}}drd\phi,
\label{eq-KDS}
\end{equation}
where $r(\phi)$ is the orbit equation of photon and we have inserted $dS=\sqrt{\text{det}\rho^{(2)}}drd\phi$.
Taking into account the above mentioned approximations, the Gaussian curvature $K_o$ and area element $dS$  can be easily obtained respectively
\begin{equation}
K_o\simeq-\frac{2M}{r^3},~~~~~~ dS\simeq r+3M.
\label{eq-kgds}
\end{equation}
Thus the surface integral of Gaussian curvature is obtained as
\begin{equation}
\begin{aligned}
-\int\int_\mathcal{D}K_odS=\int^{\phi_R}_{\phi_S}\int^{\infty}_{r_{(\phi)}}K_o\sqrt{det\rho^{(2)}}drd\phi\simeq\frac{4M}{b}+\frac{320\pi aM\gamma^2}{3b^6}.
\label{eq-kdsjf}
\end{aligned}
\end{equation}

Secondly, for the photon {in} the equatorial plane, the geodesic curvature is
\begin{equation}
\begin{aligned}
\kappa_g=-\frac{1}{\sqrt{\rho\rho^{\theta\theta}}}\beta_{\phi,r},
\label{eq-kgfc}
\end{aligned}
\end{equation}
and recalling \eqref{eq-metrspa1}, we have
\begin{equation}
\begin{aligned}
\kappa_g\simeq-\frac{2aM}{r^3}+\frac{2a(230M\pi+140\pi r)\gamma^2}{7r^7}.
\label{eq-kgjg}
\end{aligned}
\end{equation}
In addition, the line element in the integral is given by
\begin{equation}
\begin{aligned}
d\ell=\sqrt{\rho_{rr}\left(\frac{dr}{d\phi}\right)^2+\rho_{\phi\phi}}d\phi.
\label{eq-dlfc}
\end{aligned}
\end{equation}
Thus, the path integral of geodesic curvature $\kappa_g$ is
\begin{equation}
\begin{aligned}
-\int^S_R\kappa_gd\ell=\int^{\phi_R}_{\phi_S}\kappa_g\sqrt{\rho_{rr}\left(\frac{dr}{d\phi}\right)^2+\rho_{\phi\phi}}d\phi\simeq-\frac{4aM}{b^2}+\frac{15\pi^2a\gamma^2}{b^5}+\frac{2880\pi aM\gamma^2}{7b^6}.
\label{eq-kgDLjg}
\end{aligned}
\end{equation}
Finally in terms of (\ref{eq-Arpb}), (\ref{eq-kdsjf}) and (\ref{eq-kgDLjg}), the deflection angle in weak field limit of the slowly rotating black hole in dCS gravity is given as
\begin{equation}
\begin{aligned}
\alpha\simeq\frac{4M}{b}-\frac{4aM}{b^2}+\frac{15\pi^2a\gamma^2}{b^5}+\frac{10880\pi a M\gamma^2}{21b^6}.
\label{eq-kgDLzhjg}
\end{aligned}
\end{equation}
We can easily find that when $\gamma=0$, the deflection angle of the slowly rotating  black hole in dCS gravity will return to the result of the Kerr black hole in GR. Note that the result (\ref{eq-kgDLzhjg}) is for the light in the prograde motion. However for the retrograde case, the spin parameter $a$ will be changed into $-a$.




\section{Conclusion and discussion}\label{sec:conclusion}
The dCS gravity is an appealing modified gravity theory and it is one of the most widely investigated parity violating gravity theory. In this theory, the static solution is still described by Schwarzschild metric of general relativity, but Kerr black hole is not a solution to this theory and the rotating black holes will be modified by the CS coupling term. Up to date, {the complete form of fast rotating solutions is still missing, and the only axisymmetric solution was perturbatively  obtained from Schwarzschild metric in terms of small spinning parameter.} In this paper, we focused on the slowly spinning solution in the leading order correction and analyzed the effect of CS coupling term on shadow related quantities and gravitational deflection effect in weak field limit.

We analyzed the null geodesic using the Hamilton-Jacobi method, and revisited the shadow observables of the slowly rotating black hole in dCS modified gravity which {are characterized by} the shape and distortion of the shadow appearance. We found that the slowly rotating black hole with CS correction corresponds to larger black hole shadow since both the shadow radius and area {are larger comparing to that in GR.}  As the CS coupling increases, the distortion of shadow decreases {but} the oblateness increases, meaning that the CS coupling suppresses the shadow deformation brought by the spinning of black hole. Then, using the EHT observation of supermassive black hole M87* and SgrA*, we attempted to constrain the CS parameter. The spinning parameter for M87* is not lower than 0.4 from EHT observation which could exclude all slowly rotating black hole in GR or alternative gravity theories, so it is not a good probe to detect the parameters of slowly rotating black hole. While the spinning parameter for SgrA* is inconclusive, and our result showed that the shadow diameter in slowing rotating black hole of dCS theory is consistent with the EHT observation $d_{sh}=48.7 \pm 7 \mu as$ for SgrA*, indicating that it could be a candidate {for} supermassive black hole in SgrA*. In addition, we also found that the  CS term enhances the real part of QNM frequency in eikonal limit using its connection with the circular null geodesic on equatorial plane.
Finally, we  investigated the gravitational deflection angle of light in weak field limit of slowly rotating black hole spacetime with CS correction.  It was shown that the CS could enhance the deflection for prograde photon while suppresses the deflection for retrograde photon.

It is noticed that here we partly studied the shadow observables, QNM in eikonal limit and weak gravitational lensing effect, which are observations related with the null geodesic, in a slowly rotating solution which is analytically perturbative in dCS theory. Our findings can be considered  as a complementary to the {earlier research} on shadow cast \cite{Amarilla:2010zq} and strong gravitational lensing effect \cite{Chen:2010yx}. Since the non-perturbative spinning solutions to dCS gravity have been numerically constructed, it would be interesting to explore those optical observations for general rotating black hole with CS correction. In particular,  we expect future generations of EHT observations may help to detect the CS modification from GR, and even give possible constraints on the CS coupling.

\begin{acknowledgments}
This work is partly supported by  Natural Science Foundation of Jiangsu Province under Grant No.BK20211601, Fok Ying Tung Education Foundation under Grant No.171006, the Postgraduate Research \& Practice Innovation Program of Jiangsu Province under Grant No. KYCX22\_3452 and KYCX21\_3192, and Top
Talent Support Program from Yangzhou University.
\end{acknowledgments}

\begin{appendices}
\section{Null geodesic and photon region}\label{sec:app-1}
In this appendix, we shall derive the equations of motion governing  the photon trajectory  in the slowly spinning Kerr black hole \eqref{eq-metric} in CS modified gravity. The null geodesic equation is integrable and the constants of motion of photon are the Lagrangian $\mathcal{L}=\frac{1}{2}g_{\mu \nu} \dot{x}^\mu \dot{x}^\nu$. Since the spacetime of the slowly spinning Kerr black hole is independent of $t$ and $\phi$, therefore, we can define  two {conserved} quantities, the energy $E$ and the angular momentum $L_z$, as
\begin{equation}
E:=-\frac{\partial \mathcal{L}}{\partial \dot{t}}=-g_{t \phi} \dot{\phi}-g_{t t}\dot{t}, ~~~~L_z:=\frac{\partial \mathcal{L}}{\partial \dot{\phi}}=g_{\phi \phi}\dot{\phi}+g_{t \phi}\dot{t}.
\label{eq-energy}
\end{equation}
Here the dot represents the derivative {with respect to} the affine parameter $\lambda$. The Hamilton-Jacobi equation  for the particle with mass $\mu$  is given by
\begin{equation}
\mathcal{H}=-\frac{\partial S}{\partial\lambda}=\frac{1}{2}g_{\mu \nu}\frac{\partial S}{\partial x^\mu}\frac{\partial S}{\partial x^\nu}=-\frac{1}{2}\mu^2,
\label{eq-Hamilton}
\end{equation}
where $\mathcal{H}$ and $S$ are the Hamiltonian action and Jacobian action, respectively. As in the Kerr case, the Jacobian action can be separated as
\begin{equation}
S=\frac{1}{2}\mu^2 \lambda-Et+L_z\phi+S_r(r)+S_\theta(\theta),
\label{eq-Jacobi}
\end{equation}
where $\mu$ is the mass of the particle.
Combining  \eqref{eq-Hamilton} and \eqref{eq-Jacobi}, we can obtain
\begin{equation}
2\frac{\partial S}{\partial \lambda}=\mu^2=g^{tt}E^2-2g^{t\phi}EL_z+g^{\phi\phi}L_z^2+g^{rr}\big(\frac{dS_r}{dr}\big)^2+g^{\theta\theta}\big(\frac{dS_\theta}{d\theta}\big)^2.
\label{eq-H-Jacobi}
\end{equation}
For lightlike geodesics, $\mu=0$, then we can derive the equations of motion for the photon trajectory by solving  \eqref{eq-Jacobi} and \eqref{eq-H-Jacobi}.  In the current model,  the spin parameter $a$ and CS coupling parameter $\gamma$ are both small, so we calculate to the order  $a \gamma^2$ and the equations of motion are
\begin{equation}
\begin{split}
\frac{1}{E}\dot{t}=\frac{(1-2M u)(7+a \xi u^3(-14 M+ \pi u^3(70+3M u(40+63 M u))\gamma^2))}{7(1-2M u)^2}
-\frac{a^2  M u^3(1+2M u+(1-2M u)\cos(2\theta))}{(1-2 M u)^2},
\label{eq-t-motion}
\end{split}
\end{equation}
\begin{equation}
\begin{split}
\frac{1}{E}\dot{\phi}=-\frac{u^2}{7(1-2M u)}(14 a M u(a \xi u^2-1)+a \pi u^4(70+3M u(40+63M u))\gamma^2
-7\xi(2M u-1)(a^2 u^2-1)\csc^2 \theta),
\label{eq-phi-motion}
\end{split}
\end{equation}
\begin{equation}
\begin{aligned}
\frac{1}{E}\dot{r}&=\frac{1}{E}\Big(\frac{1}{g_{rr}}\frac{dS_r}{dr}\Big)\\&=
\frac{1}{g_{rr}}\sqrt{\frac{1}{\triangle}\Big(-\frac{4a\xi M u}{1-2M u}+\frac{2a\xi \pi u^4(70+3Mu(40+63Mu))\gamma^2}{7(1-2Mu)}
+\frac{1}{u^2(1-2Mu)}-\frac{a^2u^2(4M^2-\xi^2(1-2Mu))}{(1-2Mu)^2}-\eta-\xi^2\Big)},
\label{eq-r-motion}
\end{aligned}
\end{equation}
and
\begin{equation}
\begin{split}
\frac{1}{E}\dot{\theta}=\frac{1}{E}\Big(\frac{1}{g_{\theta\theta}}\frac{dS_\theta}{d\theta}\Big)=\frac{1}{g_{\theta\theta}}\sqrt{a^2 \cos^2\theta-\xi^2\cos^2\theta+\eta},
\label{eq-theta-motion}
\end{split}
\end{equation}
where $\xi=L_z/E$, $\mu=1/r$, $\eta=Q/E^2$, $Q$ is Carter constant, and $\triangle=u^{-2}-2Mu^{-1}+a^2$.
To study the black hole shadow cast, we are interested in the photon region which is relevant to the photons with unstable circular orbits. To proceed, we redefine
\begin{eqnarray}
R(u)\equiv\triangle^2\frac{1}{E^2}\Big(\frac{dS_r}{dr}\Big)^2,~~~
\Theta(\theta)\equiv\frac{1}{E^2}\Big(\frac{dS_\theta}{d\theta}\Big)^2.\label{eq-Theta}
\end{eqnarray}
The unstable circular orbits  with radius $u=u_{p}$  require
\begin{equation}
\begin{split}
R(u_p)=0, ~~\mathrm{and}~~~ \frac{dR(u)}{du}\mid_{u=u_p}=0, ~~\mathrm{and}~~~ \frac{dR^2(u)}{du^2}\mid_{u=u_p}<0
\label{eq-orbits}
\end{split}
\end{equation}
solving which we obtain  $\xi$ and $\eta$ as
\begin{equation}
\begin{split}
\xi(u_p)=\xi_K(u_p)+\frac{\pi  u^2}{7(1-2Mu)(Mu-1)a}((1-3Mu)(140+90Mu+87M^2u^2-945M^3u^3)
\\+\frac{2u^2a^2(35+55Mu-101M^2u^2-408M^3u^3+102M^4u^4+945M^5u^5)}{1-2Mu})\gamma^2\mid_{u=u_p},
\label{eq-xi-motion}
\end{split}
\end{equation}

\begin{equation}
\begin{split}
\eta(u_p)=\eta_K(u_p)-\frac{2\pi(1-3Mu)}{7((Mu-1)^2(1-2Mu))a^2}((1-3Mu)(140+90Mu+87M^2u^2-945M^3u^3)\\
+\frac{u^2(140+20Mu-233M^2u^2-1143M^3u^3+582M^4u^4+1890M^5u^5)a^2}{1-2Mu})\gamma^2\mid_{u=u_p}.
\label{eq-eta-motion}
\end{split}
\end{equation}
Here,
\begin{eqnarray}
\xi_K(u_p)=\frac{M(1-a^2u^2)-\triangle u}{au(1-Mu)},~~~\eta_K(u_p)=\frac{4Mu^3\triangle-(1-Mu)^2}{a^2u^4(1-Mu)^2}
\end{eqnarray}
{are} the corresponding results for Kerr black hole, as  the  CS coupling vanishes. Then inserting $\xi(u_p)$ and $\eta(u_p)$  into \eqref{eq-Theta},  we can determine the photon region of the rotating black hole from $\Theta(\theta)\geq 0$. For each point in this photon  region, there exists a null geodesic staying on the sphere with  $u=u_p$, along which $\phi$ and $\theta$ are governed by the equations of motion \eqref{eq-phi-motion} and \eqref{eq-theta-motion}, respectively.

\end{appendices}


\begin{thebibliography}{112}
\expandafter\ifx\csname natexlab\endcsname\relax\def\natexlab#1{#1}\fi
\expandafter\ifx\csname bibnamefont\endcsname\relax
  \def\bibnamefont#1{#1}\fi
\expandafter\ifx\csname bibfnamefont\endcsname\relax
  \def\bibfnamefont#1{#1}\fi
\expandafter\ifx\csname citenamefont\endcsname\relax
  \def\citenamefont#1{#1}\fi
\expandafter\ifx\csname url\endcsname\relax
  \def\url#1{\texttt{#1}}\fi
\expandafter\ifx\csname urlprefix\endcsname\relax\def\urlprefix{URL }\fi
\providecommand{\bibinfo}[2]{#2}
\providecommand{\eprint}[2][]{\url{#2}}

\bibitem[{\citenamefont{Abbott et~al.}(2016)}]{LIGOScientific:2016aoc}
\bibinfo{author}{\bibfnamefont{B.~P.} \bibnamefont{Abbott}}
  \bibnamefont{et~al.} (\bibinfo{collaboration}{LIGO Scientific, Virgo}),
  \bibinfo{journal}{Phys. Rev. Lett.} \textbf{\bibinfo{volume}{116}},
  \bibinfo{pages}{061102} (\bibinfo{year}{2016}), \eprint{1602.03837}.

\bibitem[{\citenamefont{Abbott et~al.}(2019)}]{LIGOScientific:2018mvr}
\bibinfo{author}{\bibfnamefont{B.~P.} \bibnamefont{Abbott}}
  \bibnamefont{et~al.} (\bibinfo{collaboration}{LIGO Scientific, Virgo}),
  \bibinfo{journal}{Phys. Rev. X} \textbf{\bibinfo{volume}{9}},
  \bibinfo{pages}{031040} (\bibinfo{year}{2019}), \eprint{1811.12907}.

\bibitem[{\citenamefont{Abbott et~al.}(2020)}]{LIGOScientific:2020aai}
\bibinfo{author}{\bibfnamefont{B.~P.} \bibnamefont{Abbott}}
  \bibnamefont{et~al.} (\bibinfo{collaboration}{LIGO Scientific, Virgo}),
  \bibinfo{journal}{Astrophys. J. Lett.} \textbf{\bibinfo{volume}{892}},
  \bibinfo{pages}{L3} (\bibinfo{year}{2020}), \eprint{2001.01761}.

\bibitem[{\citenamefont{Akiyama
  et~al.}(2019{\natexlab{a}})}]{EventHorizonTelescope:2019dse}
\bibinfo{author}{\bibfnamefont{K.}~\bibnamefont{Akiyama}} \bibnamefont{et~al.}
  (\bibinfo{collaboration}{Event Horizon Telescope}),
  \bibinfo{journal}{Astrophys. J. Lett.} \textbf{\bibinfo{volume}{875}},
  \bibinfo{pages}{L1} (\bibinfo{year}{2019}{\natexlab{a}}),
  \eprint{1906.11238}.

\bibitem[{\citenamefont{Akiyama
  et~al.}(2019{\natexlab{b}})}]{EventHorizonTelescope:2019ggy}
\bibinfo{author}{\bibfnamefont{K.}~\bibnamefont{Akiyama}} \bibnamefont{et~al.}
  (\bibinfo{collaboration}{Event Horizon Telescope}),
  \bibinfo{journal}{Astrophys. J. Lett.} \textbf{\bibinfo{volume}{875}},
  \bibinfo{pages}{L6} (\bibinfo{year}{2019}{\natexlab{b}}),
  \eprint{1906.11243}.

\bibitem[{\citenamefont{Akiyama
  et~al.}(2019{\natexlab{c}})}]{EventHorizonTelescope:2019ths}
\bibinfo{author}{\bibfnamefont{K.}~\bibnamefont{Akiyama}} \bibnamefont{et~al.}
  (\bibinfo{collaboration}{Event Horizon Telescope}),
  \bibinfo{journal}{Astrophys. J. Lett.} \textbf{\bibinfo{volume}{875}},
  \bibinfo{pages}{L4} (\bibinfo{year}{2019}{\natexlab{c}}),
  \eprint{1906.11241}.

\bibitem[{\citenamefont{Akiyama
  et~al.}(2022{\natexlab{a}})}]{EventHorizonTelescope:2022xnr}
\bibinfo{author}{\bibfnamefont{K.}~\bibnamefont{Akiyama}} \bibnamefont{et~al.}
  (\bibinfo{collaboration}{Event Horizon Telescope}),
  \bibinfo{journal}{Astrophys. J. Lett.} \textbf{\bibinfo{volume}{930}},
  \bibinfo{pages}{L12} (\bibinfo{year}{2022}{\natexlab{a}}).

\bibitem[{\citenamefont{Akiyama
  et~al.}(2022{\natexlab{b}})}]{EventHorizonTelescope:2022xqj}
\bibinfo{author}{\bibfnamefont{K.}~\bibnamefont{Akiyama}} \bibnamefont{et~al.}
  (\bibinfo{collaboration}{Event Horizon Telescope}),
  \bibinfo{journal}{Astrophys. J. Lett.} \textbf{\bibinfo{volume}{930}},
  \bibinfo{pages}{L17} (\bibinfo{year}{2022}{\natexlab{b}}).

\bibitem[{\citenamefont{Synge}(1966)}]{Synge:1966okc}
\bibinfo{author}{\bibfnamefont{J.~L.} \bibnamefont{Synge}},
  \bibinfo{journal}{Mon. Not. Roy. Astron. Soc.}
  \textbf{\bibinfo{volume}{131}}, \bibinfo{pages}{463} (\bibinfo{year}{1966}).

\bibitem[{\citenamefont{Luminet}(1979)}]{Luminet:1979nyg}
\bibinfo{author}{\bibfnamefont{J.~P.} \bibnamefont{Luminet}},
  \bibinfo{journal}{Astron. Astrophys.} \textbf{\bibinfo{volume}{75}},
  \bibinfo{pages}{228} (\bibinfo{year}{1979}).

\bibitem[{\citenamefont{Bardeen}(1973)}]{bardeen1973houches}
\bibinfo{author}{\bibfnamefont{J.}~\bibnamefont{Bardeen}},
  \emph{\bibinfo{title}{Les houches summer school of theoretical physics: Black
  holes}} (\bibinfo{year}{1973}).

\bibitem[{\citenamefont{Falcke et~al.}(2000)\citenamefont{Falcke, Melia, and
  Agol}}]{Falcke:1999pj}
\bibinfo{author}{\bibfnamefont{H.}~\bibnamefont{Falcke}},
  \bibinfo{author}{\bibfnamefont{F.}~\bibnamefont{Melia}}, \bibnamefont{and}
  \bibinfo{author}{\bibfnamefont{E.}~\bibnamefont{Agol}},
  \bibinfo{journal}{Astrophys. J. Lett.} \textbf{\bibinfo{volume}{528}},
  \bibinfo{pages}{L13} (\bibinfo{year}{2000}), \eprint{astro-ph/9912263}.

\bibitem[{\citenamefont{Virbhadra and Ellis}(2000)}]{Virbhadra:1999nm}
\bibinfo{author}{\bibfnamefont{K.~S.} \bibnamefont{Virbhadra}}
  \bibnamefont{and} \bibinfo{author}{\bibfnamefont{G.~F.~R.}
  \bibnamefont{Ellis}}, \bibinfo{journal}{Phys. Rev. D}
  \textbf{\bibinfo{volume}{62}}, \bibinfo{pages}{084003}
  (\bibinfo{year}{2000}), \eprint{astro-ph/9904193}.

\bibitem[{\citenamefont{Shen et~al.}(2005)\citenamefont{Shen, Lo, Liang, Ho,
  and Zhao}}]{Shen:2005cw}
\bibinfo{author}{\bibfnamefont{Z.-Q.} \bibnamefont{Shen}},
  \bibinfo{author}{\bibfnamefont{K.~Y.} \bibnamefont{Lo}},
  \bibinfo{author}{\bibfnamefont{M.~C.} \bibnamefont{Liang}},
  \bibinfo{author}{\bibfnamefont{P.~T.~P.} \bibnamefont{Ho}}, \bibnamefont{and}
  \bibinfo{author}{\bibfnamefont{J.~H.} \bibnamefont{Zhao}},
  \bibinfo{journal}{Nature} \textbf{\bibinfo{volume}{438}}, \bibinfo{pages}{62}
  (\bibinfo{year}{2005}), \eprint{astro-ph/0512515}.

\bibitem[{\citenamefont{Younsi et~al.}(2016)\citenamefont{Younsi, Zhidenko,
  Rezzolla, Konoplya, and Mizuno}}]{Younsi:2016azx}
\bibinfo{author}{\bibfnamefont{Z.}~\bibnamefont{Younsi}},
  \bibinfo{author}{\bibfnamefont{A.}~\bibnamefont{Zhidenko}},
  \bibinfo{author}{\bibfnamefont{L.}~\bibnamefont{Rezzolla}},
  \bibinfo{author}{\bibfnamefont{R.}~\bibnamefont{Konoplya}}, \bibnamefont{and}
  \bibinfo{author}{\bibfnamefont{Y.}~\bibnamefont{Mizuno}},
  \bibinfo{journal}{Phys. Rev. D} \textbf{\bibinfo{volume}{94}},
  \bibinfo{pages}{084025} (\bibinfo{year}{2016}), \eprint{1607.05767}.

\bibitem[{\citenamefont{Atamurotov et~al.}(2013)\citenamefont{Atamurotov,
  Abdujabbarov, and Ahmedov}}]{Atamurotov:2013sca}
\bibinfo{author}{\bibfnamefont{F.}~\bibnamefont{Atamurotov}},
  \bibinfo{author}{\bibfnamefont{A.}~\bibnamefont{Abdujabbarov}},
  \bibnamefont{and} \bibinfo{author}{\bibfnamefont{B.}~\bibnamefont{Ahmedov}},
  \bibinfo{journal}{Phys. Rev. D} \textbf{\bibinfo{volume}{88}},
  \bibinfo{pages}{064004} (\bibinfo{year}{2013}).

\bibitem[{\citenamefont{Atamurotov et~al.}(2016)\citenamefont{Atamurotov,
  Ghosh, and Ahmedov}}]{Atamurotov:2015xfa}
\bibinfo{author}{\bibfnamefont{F.}~\bibnamefont{Atamurotov}},
  \bibinfo{author}{\bibfnamefont{S.~G.} \bibnamefont{Ghosh}}, \bibnamefont{and}
  \bibinfo{author}{\bibfnamefont{B.}~\bibnamefont{Ahmedov}},
  \bibinfo{journal}{Eur. Phys. J. C} \textbf{\bibinfo{volume}{76}},
  \bibinfo{pages}{273} (\bibinfo{year}{2016}), \eprint{1506.03690}.

\bibitem[{\citenamefont{Amir et~al.}(2018)\citenamefont{Amir, Singh, and
  Ghosh}}]{Amir:2017slq}
\bibinfo{author}{\bibfnamefont{M.}~\bibnamefont{Amir}},
  \bibinfo{author}{\bibfnamefont{B.~P.} \bibnamefont{Singh}}, \bibnamefont{and}
  \bibinfo{author}{\bibfnamefont{S.~G.} \bibnamefont{Ghosh}},
  \bibinfo{journal}{Eur. Phys. J. C} \textbf{\bibinfo{volume}{78}},
  \bibinfo{pages}{399} (\bibinfo{year}{2018}), \eprint{1707.09521}.

\bibitem[{\citenamefont{Eiroa and Sendra}(2018)}]{Eiroa:2017uuq}
\bibinfo{author}{\bibfnamefont{E.~F.} \bibnamefont{Eiroa}} \bibnamefont{and}
  \bibinfo{author}{\bibfnamefont{C.~M.} \bibnamefont{Sendra}},
  \bibinfo{journal}{Eur. Phys. J. C} \textbf{\bibinfo{volume}{78}},
  \bibinfo{pages}{91} (\bibinfo{year}{2018}), \eprint{1711.08380}.

\bibitem[{\citenamefont{Vagnozzi and Visinelli}(2019)}]{Vagnozzi:2019apd}
\bibinfo{author}{\bibfnamefont{S.}~\bibnamefont{Vagnozzi}} \bibnamefont{and}
  \bibinfo{author}{\bibfnamefont{L.}~\bibnamefont{Visinelli}},
  \bibinfo{journal}{Phys. Rev. D} \textbf{\bibinfo{volume}{100}},
  \bibinfo{pages}{024020} (\bibinfo{year}{2019}), \eprint{1905.12421}.

\bibitem[{\citenamefont{Long et~al.}(2020)\citenamefont{Long, Chen, Wang, and
  Jing}}]{Long:2020wqj}
\bibinfo{author}{\bibfnamefont{F.}~\bibnamefont{Long}},
  \bibinfo{author}{\bibfnamefont{S.}~\bibnamefont{Chen}},
  \bibinfo{author}{\bibfnamefont{M.}~\bibnamefont{Wang}}, \bibnamefont{and}
  \bibinfo{author}{\bibfnamefont{J.}~\bibnamefont{Jing}},
  \bibinfo{journal}{Eur. Phys. J. C} \textbf{\bibinfo{volume}{80}},
  \bibinfo{pages}{1180} (\bibinfo{year}{2020}), \eprint{2009.07508}.

\bibitem[{\citenamefont{Banerjee et~al.}(2020)\citenamefont{Banerjee,
  Chakraborty, and SenGupta}}]{Banerjee:2019nnj}
\bibinfo{author}{\bibfnamefont{I.}~\bibnamefont{Banerjee}},
  \bibinfo{author}{\bibfnamefont{S.}~\bibnamefont{Chakraborty}},
  \bibnamefont{and} \bibinfo{author}{\bibfnamefont{S.}~\bibnamefont{SenGupta}},
  \bibinfo{journal}{Phys. Rev. D} \textbf{\bibinfo{volume}{101}},
  \bibinfo{pages}{041301} (\bibinfo{year}{2020}), \eprint{1909.09385}.

\bibitem[{\citenamefont{Kumar et~al.}(2020{\natexlab{a}})\citenamefont{Kumar,
  Ghosh, and Wang}}]{Kumar:2020hgm}
\bibinfo{author}{\bibfnamefont{R.}~\bibnamefont{Kumar}},
  \bibinfo{author}{\bibfnamefont{S.~G.} \bibnamefont{Ghosh}}, \bibnamefont{and}
  \bibinfo{author}{\bibfnamefont{A.}~\bibnamefont{Wang}},
  \bibinfo{journal}{Phys. Rev. D} \textbf{\bibinfo{volume}{101}},
  \bibinfo{pages}{104001} (\bibinfo{year}{2020}{\natexlab{a}}),
  \eprint{2001.00460}.

\bibitem[{\citenamefont{Qian et~al.}(2022)\citenamefont{Qian, Chen, Shao, Wang,
  and Yue}}]{Qian:2021qow}
\bibinfo{author}{\bibfnamefont{W.-L.} \bibnamefont{Qian}},
  \bibinfo{author}{\bibfnamefont{S.}~\bibnamefont{Chen}},
  \bibinfo{author}{\bibfnamefont{C.-G.} \bibnamefont{Shao}},
  \bibinfo{author}{\bibfnamefont{B.}~\bibnamefont{Wang}}, \bibnamefont{and}
  \bibinfo{author}{\bibfnamefont{R.-H.} \bibnamefont{Yue}},
  \bibinfo{journal}{Eur. Phys. J. C} \textbf{\bibinfo{volume}{82}},
  \bibinfo{pages}{91} (\bibinfo{year}{2022}), \eprint{2102.03820}.

\bibitem[{\citenamefont{Zeng et~al.}(2022)\citenamefont{Zeng, He, and
  Li}}]{Zeng:2021mok}
\bibinfo{author}{\bibfnamefont{X.-X.} \bibnamefont{Zeng}},
  \bibinfo{author}{\bibfnamefont{K.-J.} \bibnamefont{He}}, \bibnamefont{and}
  \bibinfo{author}{\bibfnamefont{G.-P.} \bibnamefont{Li}},
  \bibinfo{journal}{Sci. China Phys. Mech. Astron.}
  \textbf{\bibinfo{volume}{65}}, \bibinfo{pages}{290411}
  (\bibinfo{year}{2022}), \eprint{2111.05090}.

\bibitem[{\citenamefont{Lin et~al.}(2022)\citenamefont{Lin, Patel, and
  Pu}}]{Lin:2022ksb}
\bibinfo{author}{\bibfnamefont{F.-L.} \bibnamefont{Lin}},
  \bibinfo{author}{\bibfnamefont{A.}~\bibnamefont{Patel}}, \bibnamefont{and}
  \bibinfo{author}{\bibfnamefont{H.-Y.} \bibnamefont{Pu}}
  (\bibinfo{year}{2022}), \eprint{2202.13559}.

\bibitem[{\citenamefont{Sun et~al.}(2022)\citenamefont{Sun, Liu, Qian, and
  Yue}}]{Sun:2022wya}
\bibinfo{author}{\bibfnamefont{C.}~\bibnamefont{Sun}},
  \bibinfo{author}{\bibfnamefont{Y.}~\bibnamefont{Liu}},
  \bibinfo{author}{\bibfnamefont{W.-L.} \bibnamefont{Qian}}, \bibnamefont{and}
  \bibinfo{author}{\bibfnamefont{R.}~\bibnamefont{Yue}},
  \bibinfo{journal}{Chin. Phys. C} \textbf{\bibinfo{volume}{46}},
  \bibinfo{pages}{065103} (\bibinfo{year}{2022}), \eprint{2201.01890}.

\bibitem[{\citenamefont{\c{C}imdiker et~al.}(2021)\citenamefont{\c{C}imdiker,
  Demir, and \"Ovg\"un}}]{Cimdiker:2021cpz}
\bibinfo{author}{\bibfnamefont{I.}~\bibnamefont{\c{C}imdiker}},
  \bibinfo{author}{\bibfnamefont{D.}~\bibnamefont{Demir}}, \bibnamefont{and}
  \bibinfo{author}{\bibfnamefont{A.}~\bibnamefont{\"Ovg\"un}},
  \bibinfo{journal}{Phys. Dark Univ.} \textbf{\bibinfo{volume}{34}},
  \bibinfo{pages}{100900} (\bibinfo{year}{2021}), \eprint{2110.11904}.

\bibitem[{\citenamefont{Zhong et~al.}(2021)\citenamefont{Zhong, Hu, Yan, Guo,
  and Chen}}]{Zhong:2021mty}
\bibinfo{author}{\bibfnamefont{Z.}~\bibnamefont{Zhong}},
  \bibinfo{author}{\bibfnamefont{Z.}~\bibnamefont{Hu}},
  \bibinfo{author}{\bibfnamefont{H.}~\bibnamefont{Yan}},
  \bibinfo{author}{\bibfnamefont{M.}~\bibnamefont{Guo}}, \bibnamefont{and}
  \bibinfo{author}{\bibfnamefont{B.}~\bibnamefont{Chen}},
  \bibinfo{journal}{Phys. Rev. D} \textbf{\bibinfo{volume}{104}},
  \bibinfo{pages}{104028} (\bibinfo{year}{2021}), \eprint{2108.06140}.

\bibitem[{\citenamefont{Hou et~al.}(2021)\citenamefont{Hou, Guo, and
  Chen}}]{Hou:2021okc}
\bibinfo{author}{\bibfnamefont{Y.}~\bibnamefont{Hou}},
  \bibinfo{author}{\bibfnamefont{M.}~\bibnamefont{Guo}}, \bibnamefont{and}
  \bibinfo{author}{\bibfnamefont{B.}~\bibnamefont{Chen}},
  \bibinfo{journal}{Phys. Rev. D} \textbf{\bibinfo{volume}{104}},
  \bibinfo{pages}{024001} (\bibinfo{year}{2021}), \eprint{2103.04369}.

\bibitem[{\citenamefont{Gan et~al.}(2021)\citenamefont{Gan, Wang, Wu, and
  Yang}}]{Gan:2021pwu}
\bibinfo{author}{\bibfnamefont{Q.}~\bibnamefont{Gan}},
  \bibinfo{author}{\bibfnamefont{P.}~\bibnamefont{Wang}},
  \bibinfo{author}{\bibfnamefont{H.}~\bibnamefont{Wu}}, \bibnamefont{and}
  \bibinfo{author}{\bibfnamefont{H.}~\bibnamefont{Yang}},
  \bibinfo{journal}{Phys. Rev. D} \textbf{\bibinfo{volume}{104}},
  \bibinfo{pages}{024003} (\bibinfo{year}{2021}), \eprint{2104.08703}.

\bibitem[{\citenamefont{Chang and Zhu}(2021)}]{Chang:2021ngy}
\bibinfo{author}{\bibfnamefont{Z.}~\bibnamefont{Chang}} \bibnamefont{and}
  \bibinfo{author}{\bibfnamefont{Q.-H.} \bibnamefont{Zhu}},
  \bibinfo{journal}{JCAP} \textbf{\bibinfo{volume}{09}}, \bibinfo{pages}{003}
  (\bibinfo{year}{2021}), \eprint{2104.14221}.

\bibitem[{\citenamefont{Wang et~al.}(2021)\citenamefont{Wang, Chen, and
  Jing}}]{Wang:2021ara}
\bibinfo{author}{\bibfnamefont{M.}~\bibnamefont{Wang}},
  \bibinfo{author}{\bibfnamefont{S.}~\bibnamefont{Chen}}, \bibnamefont{and}
  \bibinfo{author}{\bibfnamefont{J.}~\bibnamefont{Jing}},
  \bibinfo{journal}{Phys. Rev. D} \textbf{\bibinfo{volume}{104}},
  \bibinfo{pages}{084021} (\bibinfo{year}{2021}), \eprint{2104.12304}.

\bibitem[{\citenamefont{Shaikh et~al.}(2021)\citenamefont{Shaikh, Paul,
  Banerjee, and Sarkar}}]{Shaikh:2021cvl}
\bibinfo{author}{\bibfnamefont{R.}~\bibnamefont{Shaikh}},
  \bibinfo{author}{\bibfnamefont{S.}~\bibnamefont{Paul}},
  \bibinfo{author}{\bibfnamefont{P.}~\bibnamefont{Banerjee}}, \bibnamefont{and}
  \bibinfo{author}{\bibfnamefont{T.}~\bibnamefont{Sarkar}}
  (\bibinfo{year}{2021}), \eprint{2105.12057}.

\bibitem[{\citenamefont{Guo et~al.}(2020)\citenamefont{Guo, Liu, Kuang, and
  Wang}}]{Guo:2020blq}
\bibinfo{author}{\bibfnamefont{H.}~\bibnamefont{Guo}},
  \bibinfo{author}{\bibfnamefont{H.}~\bibnamefont{Liu}},
  \bibinfo{author}{\bibfnamefont{X.-M.} \bibnamefont{Kuang}}, \bibnamefont{and}
  \bibinfo{author}{\bibfnamefont{B.}~\bibnamefont{Wang}},
  \bibinfo{journal}{Phys. Rev. D} \textbf{\bibinfo{volume}{102}},
  \bibinfo{pages}{124019} (\bibinfo{year}{2020}), \eprint{2007.04197}.

\bibitem[{\citenamefont{Meng et~al.}(2022)\citenamefont{Meng, Kuang, and
  Tang}}]{Meng:2022kjs}
\bibinfo{author}{\bibfnamefont{Y.}~\bibnamefont{Meng}},
  \bibinfo{author}{\bibfnamefont{X.-M.} \bibnamefont{Kuang}}, \bibnamefont{and}
  \bibinfo{author}{\bibfnamefont{Z.-Y.} \bibnamefont{Tang}},
  \bibinfo{journal}{Phys. Rev. D} \textbf{\bibinfo{volume}{106}},
  \bibinfo{pages}{064006} (\bibinfo{year}{2022}), \eprint{2204.00897}.

\bibitem[{\citenamefont{Kuang et~al.}(2022)\citenamefont{Kuang, Tang, Wang, and
  Wang}}]{Kuang:2022ojj}
\bibinfo{author}{\bibfnamefont{X.-M.} \bibnamefont{Kuang}},
  \bibinfo{author}{\bibfnamefont{Z.-Y.} \bibnamefont{Tang}},
  \bibinfo{author}{\bibfnamefont{B.}~\bibnamefont{Wang}}, \bibnamefont{and}
  \bibinfo{author}{\bibfnamefont{A.}~\bibnamefont{Wang}},
  \bibinfo{journal}{Phys. Rev. D} \textbf{\bibinfo{volume}{106}},
  \bibinfo{pages}{064012} (\bibinfo{year}{2022}), \eprint{2206.05878}.

\bibitem[{\citenamefont{Kuang and \"Ovg\"un}(2022)}]{Kuang:2022xjp}
\bibinfo{author}{\bibfnamefont{X.-M.} \bibnamefont{Kuang}} \bibnamefont{and}
  \bibinfo{author}{\bibfnamefont{A.}~\bibnamefont{\"Ovg\"un}},
  \bibinfo{journal}{Annals Phys.} \textbf{\bibinfo{volume}{447}},
  \bibinfo{pages}{169147} (\bibinfo{year}{2022}), \eprint{2205.11003}.

\bibitem[{\citenamefont{Tang et~al.}(2022)\citenamefont{Tang, Kuang, Wang, and
  Qian}}]{Tang:2022hsu}
\bibinfo{author}{\bibfnamefont{Z.-Y.} \bibnamefont{Tang}},
  \bibinfo{author}{\bibfnamefont{X.-M.} \bibnamefont{Kuang}},
  \bibinfo{author}{\bibfnamefont{B.}~\bibnamefont{Wang}}, \bibnamefont{and}
  \bibinfo{author}{\bibfnamefont{W.-L.} \bibnamefont{Qian}},
  \bibinfo{journal}{Sci. Bull.} \textbf{\bibinfo{volume}{67}},
  \bibinfo{pages}{2272} (\bibinfo{year}{2022}), \eprint{2206.08608}.

\bibitem[{\citenamefont{Hioki and Maeda}(2009)}]{Hioki:2009na}
\bibinfo{author}{\bibfnamefont{K.}~\bibnamefont{Hioki}} \bibnamefont{and}
  \bibinfo{author}{\bibfnamefont{K.-i.} \bibnamefont{Maeda}},
  \bibinfo{journal}{Phys. Rev. D} \textbf{\bibinfo{volume}{80}},
  \bibinfo{pages}{024042} (\bibinfo{year}{2009}), \eprint{0904.3575}.

\bibitem[{\citenamefont{Kramer et~al.}(2004)\citenamefont{Kramer, Backer,
  Cordes, Lazio, Stappers, and Johnston}}]{Kramer:2004hd}
\bibinfo{author}{\bibfnamefont{M.}~\bibnamefont{Kramer}},
  \bibinfo{author}{\bibfnamefont{D.~C.} \bibnamefont{Backer}},
  \bibinfo{author}{\bibfnamefont{J.~M.} \bibnamefont{Cordes}},
  \bibinfo{author}{\bibfnamefont{T.~J.~W.} \bibnamefont{Lazio}},
  \bibinfo{author}{\bibfnamefont{B.~W.} \bibnamefont{Stappers}},
  \bibnamefont{and} \bibinfo{author}{\bibfnamefont{S.}~\bibnamefont{Johnston}},
  \bibinfo{journal}{New Astron. Rev.} \textbf{\bibinfo{volume}{48}},
  \bibinfo{pages}{993} (\bibinfo{year}{2004}), \eprint{astro-ph/0409379}.

\bibitem[{\citenamefont{Tsupko}(2017)}]{Tsupko:2017rdo}
\bibinfo{author}{\bibfnamefont{O.~Y.} \bibnamefont{Tsupko}},
  \bibinfo{journal}{Phys. Rev. D} \textbf{\bibinfo{volume}{95}},
  \bibinfo{pages}{104058} (\bibinfo{year}{2017}), \eprint{1702.04005}.

\bibitem[{\citenamefont{Cunha et~al.}(2019{\natexlab{a}})\citenamefont{Cunha,
  Herdeiro, and Radu}}]{Cunha:2019ikd}
\bibinfo{author}{\bibfnamefont{P.~V.~P.} \bibnamefont{Cunha}},
  \bibinfo{author}{\bibfnamefont{C.~A.~R.} \bibnamefont{Herdeiro}},
  \bibnamefont{and} \bibinfo{author}{\bibfnamefont{E.}~\bibnamefont{Radu}},
  \bibinfo{journal}{Universe} \textbf{\bibinfo{volume}{5}},
  \bibinfo{pages}{220} (\bibinfo{year}{2019}{\natexlab{a}}),
  \eprint{1909.08039}.

\bibitem[{\citenamefont{Cunha et~al.}(2019{\natexlab{b}})\citenamefont{Cunha,
  Herdeiro, and Radu}}]{Cunha:2019dwb}
\bibinfo{author}{\bibfnamefont{P.~V.~P.} \bibnamefont{Cunha}},
  \bibinfo{author}{\bibfnamefont{C.~A.~R.} \bibnamefont{Herdeiro}},
  \bibnamefont{and} \bibinfo{author}{\bibfnamefont{E.}~\bibnamefont{Radu}},
  \bibinfo{journal}{Phys. Rev. Lett.} \textbf{\bibinfo{volume}{123}},
  \bibinfo{pages}{011101} (\bibinfo{year}{2019}{\natexlab{b}}),
  \eprint{1904.09997}.

\bibitem[{\citenamefont{Kumar and Ghosh}(2020{\natexlab{a}})}]{Kumar:2018ple}
\bibinfo{author}{\bibfnamefont{R.}~\bibnamefont{Kumar}} \bibnamefont{and}
  \bibinfo{author}{\bibfnamefont{S.~G.} \bibnamefont{Ghosh}},
  \bibinfo{journal}{Astrophys. J.} \textbf{\bibinfo{volume}{892}},
  \bibinfo{pages}{78} (\bibinfo{year}{2020}{\natexlab{a}}),
  \eprint{1811.01260}.

\bibitem[{\citenamefont{Khodadi et~al.}(2020)\citenamefont{Khodadi, Allahyari,
  Vagnozzi, and Mota}}]{Khodadi:2020jij}
\bibinfo{author}{\bibfnamefont{M.}~\bibnamefont{Khodadi}},
  \bibinfo{author}{\bibfnamefont{A.}~\bibnamefont{Allahyari}},
  \bibinfo{author}{\bibfnamefont{S.}~\bibnamefont{Vagnozzi}}, \bibnamefont{and}
  \bibinfo{author}{\bibfnamefont{D.~F.} \bibnamefont{Mota}},
  \bibinfo{journal}{JCAP} \textbf{\bibinfo{volume}{09}}, \bibinfo{pages}{026}
  (\bibinfo{year}{2020}), \eprint{2005.05992}.

\bibitem[{\citenamefont{Bad\'\i{}a and Eiroa}(2021)}]{Badia:2021kpk}
\bibinfo{author}{\bibfnamefont{J.}~\bibnamefont{Bad\'\i{}a}} \bibnamefont{and}
  \bibinfo{author}{\bibfnamefont{E.~F.} \bibnamefont{Eiroa}},
  \bibinfo{journal}{Phys. Rev. D} \textbf{\bibinfo{volume}{104}},
  \bibinfo{pages}{084055} (\bibinfo{year}{2021}), \eprint{2106.07601}.

\bibitem[{\citenamefont{Wei and Liu}(2013)}]{Wei:2013kza}
\bibinfo{author}{\bibfnamefont{S.-W.} \bibnamefont{Wei}} \bibnamefont{and}
  \bibinfo{author}{\bibfnamefont{Y.-X.} \bibnamefont{Liu}},
  \bibinfo{journal}{JCAP} \textbf{\bibinfo{volume}{11}}, \bibinfo{pages}{063}
  (\bibinfo{year}{2013}), \eprint{1311.4251}.

\bibitem[{\citenamefont{Allahyari et~al.}(2020)\citenamefont{Allahyari,
  Khodadi, Vagnozzi, and Mota}}]{Allahyari:2019jqz}
\bibinfo{author}{\bibfnamefont{A.}~\bibnamefont{Allahyari}},
  \bibinfo{author}{\bibfnamefont{M.}~\bibnamefont{Khodadi}},
  \bibinfo{author}{\bibfnamefont{S.}~\bibnamefont{Vagnozzi}}, \bibnamefont{and}
  \bibinfo{author}{\bibfnamefont{D.~F.} \bibnamefont{Mota}},
  \bibinfo{journal}{JCAP} \textbf{\bibinfo{volume}{02}}, \bibinfo{pages}{003}
  (\bibinfo{year}{2020}), \eprint{1912.08231}.

\bibitem[{\citenamefont{Kumar and Ghosh}(2020{\natexlab{b}})}]{Kumar:2020owy}
\bibinfo{author}{\bibfnamefont{R.}~\bibnamefont{Kumar}} \bibnamefont{and}
  \bibinfo{author}{\bibfnamefont{S.~G.} \bibnamefont{Ghosh}},
  \bibinfo{journal}{JCAP} \textbf{\bibinfo{volume}{07}}, \bibinfo{pages}{053}
  (\bibinfo{year}{2020}{\natexlab{b}}), \eprint{2003.08927}.

\bibitem[{\citenamefont{Belhaj et~al.}(2021)\citenamefont{Belhaj, Benali,
  Balali, Hadri, and El~Moumni}}]{Belhaj:2020kwv}
\bibinfo{author}{\bibfnamefont{A.}~\bibnamefont{Belhaj}},
  \bibinfo{author}{\bibfnamefont{M.}~\bibnamefont{Benali}},
  \bibinfo{author}{\bibfnamefont{A.~E.} \bibnamefont{Balali}},
  \bibinfo{author}{\bibfnamefont{W.~E.} \bibnamefont{Hadri}}, \bibnamefont{and}
  \bibinfo{author}{\bibfnamefont{H.}~\bibnamefont{El~Moumni}},
  \bibinfo{journal}{Int. J. Geom. Meth. Mod. Phys.}
  \textbf{\bibinfo{volume}{18}}, \bibinfo{pages}{2150188}
  (\bibinfo{year}{2021}), \eprint{2007.09058}.

\bibitem[{\citenamefont{Cardoso et~al.}(2009)\citenamefont{Cardoso, Miranda,
  Berti, Witek, and Zanchin}}]{Cardoso:2008bp}
\bibinfo{author}{\bibfnamefont{V.}~\bibnamefont{Cardoso}},
  \bibinfo{author}{\bibfnamefont{A.~S.} \bibnamefont{Miranda}},
  \bibinfo{author}{\bibfnamefont{E.}~\bibnamefont{Berti}},
  \bibinfo{author}{\bibfnamefont{H.}~\bibnamefont{Witek}}, \bibnamefont{and}
  \bibinfo{author}{\bibfnamefont{V.~T.} \bibnamefont{Zanchin}},
  \bibinfo{journal}{Phys. Rev. D} \textbf{\bibinfo{volume}{79}},
  \bibinfo{pages}{064016} (\bibinfo{year}{2009}), \eprint{0812.1806}.

\bibitem[{\citenamefont{Jusufi}(2020{\natexlab{a}})}]{Jusufi:2019ltj}
\bibinfo{author}{\bibfnamefont{K.}~\bibnamefont{Jusufi}},
  \bibinfo{journal}{Phys. Rev. D} \textbf{\bibinfo{volume}{101}},
  \bibinfo{pages}{084055} (\bibinfo{year}{2020}{\natexlab{a}}),
  \eprint{1912.13320}.

\bibitem[{\citenamefont{Liu et~al.}(2020)\citenamefont{Liu, Zhu, Wu, Jusufi,
  Jamil, Azreg-A\"\i{}nou, and Wang}}]{Liu:2020ola}
\bibinfo{author}{\bibfnamefont{C.}~\bibnamefont{Liu}},
  \bibinfo{author}{\bibfnamefont{T.}~\bibnamefont{Zhu}},
  \bibinfo{author}{\bibfnamefont{Q.}~\bibnamefont{Wu}},
  \bibinfo{author}{\bibfnamefont{K.}~\bibnamefont{Jusufi}},
  \bibinfo{author}{\bibfnamefont{M.}~\bibnamefont{Jamil}},
  \bibinfo{author}{\bibfnamefont{M.}~\bibnamefont{Azreg-A\"\i{}nou}},
  \bibnamefont{and} \bibinfo{author}{\bibfnamefont{A.}~\bibnamefont{Wang}},
  \bibinfo{journal}{Phys. Rev. D} \textbf{\bibinfo{volume}{101}},
  \bibinfo{pages}{084001} (\bibinfo{year}{2020}), \bibinfo{note}{[Erratum:
  Phys.Rev.D 103, 089902 (2021)]}, \eprint{2003.00477}.

\bibitem[{\citenamefont{Jusufi}(2020{\natexlab{b}})}]{Jusufi:2020dhz}
\bibinfo{author}{\bibfnamefont{K.}~\bibnamefont{Jusufi}},
  \bibinfo{journal}{Phys. Rev. D} \textbf{\bibinfo{volume}{101}},
  \bibinfo{pages}{124063} (\bibinfo{year}{2020}{\natexlab{b}}),
  \eprint{2004.04664}.

\bibitem[{\citenamefont{Jackiw and Pi}(2003)}]{Jackiw:2003pm}
\bibinfo{author}{\bibfnamefont{R.}~\bibnamefont{Jackiw}} \bibnamefont{and}
  \bibinfo{author}{\bibfnamefont{S.~Y.} \bibnamefont{Pi}},
  \bibinfo{journal}{Phys. Rev. D} \textbf{\bibinfo{volume}{68}},
  \bibinfo{pages}{104012} (\bibinfo{year}{2003}), \eprint{gr-qc/0308071}.

\bibitem[{\citenamefont{Alexander and Yunes}(2009)}]{Alexander:2009tp}
\bibinfo{author}{\bibfnamefont{S.}~\bibnamefont{Alexander}} \bibnamefont{and}
  \bibinfo{author}{\bibfnamefont{N.}~\bibnamefont{Yunes}},
  \bibinfo{journal}{Phys. Rept.} \textbf{\bibinfo{volume}{480}},
  \bibinfo{pages}{1} (\bibinfo{year}{2009}), \eprint{0907.2562}.

\bibitem[{\citenamefont{Alexander et~al.}(2008)\citenamefont{Alexander, Finn,
  and Yunes}}]{Alexander:2007kv}
\bibinfo{author}{\bibfnamefont{S.}~\bibnamefont{Alexander}},
  \bibinfo{author}{\bibfnamefont{L.~S.} \bibnamefont{Finn}}, \bibnamefont{and}
  \bibinfo{author}{\bibfnamefont{N.}~\bibnamefont{Yunes}},
  \bibinfo{journal}{Phys. Rev. D} \textbf{\bibinfo{volume}{78}},
  \bibinfo{pages}{066005} (\bibinfo{year}{2008}), \eprint{0712.2542}.

\bibitem[{\citenamefont{Alexander and Yunes}(2007)}]{Alexander:2007zg}
\bibinfo{author}{\bibfnamefont{S.}~\bibnamefont{Alexander}} \bibnamefont{and}
  \bibinfo{author}{\bibfnamefont{N.}~\bibnamefont{Yunes}},
  \bibinfo{journal}{Phys. Rev. Lett.} \textbf{\bibinfo{volume}{99}},
  \bibinfo{pages}{241101} (\bibinfo{year}{2007}), \eprint{hep-th/0703265}.

\bibitem[{\citenamefont{Smith et~al.}(2008)\citenamefont{Smith, Erickcek,
  Caldwell, and Kamionkowski}}]{Smith:2007jm}
\bibinfo{author}{\bibfnamefont{T.~L.} \bibnamefont{Smith}},
  \bibinfo{author}{\bibfnamefont{A.~L.} \bibnamefont{Erickcek}},
  \bibinfo{author}{\bibfnamefont{R.~R.} \bibnamefont{Caldwell}},
  \bibnamefont{and}
  \bibinfo{author}{\bibfnamefont{M.}~\bibnamefont{Kamionkowski}},
  \bibinfo{journal}{Phys. Rev. D} \textbf{\bibinfo{volume}{77}},
  \bibinfo{pages}{024015} (\bibinfo{year}{2008}), \eprint{0708.0001}.

\bibitem[{\citenamefont{Yunes and Pretorius}(2009)}]{Yunes:2009hc}
\bibinfo{author}{\bibfnamefont{N.}~\bibnamefont{Yunes}} \bibnamefont{and}
  \bibinfo{author}{\bibfnamefont{F.}~\bibnamefont{Pretorius}},
  \bibinfo{journal}{Phys. Rev. D} \textbf{\bibinfo{volume}{79}},
  \bibinfo{pages}{084043} (\bibinfo{year}{2009}), \eprint{0902.4669}.

\bibitem[{\citenamefont{Yunes et~al.}(2010)\citenamefont{Yunes, Psaltis, Ozel,
  and Loeb}}]{Yunes:2009ch}
\bibinfo{author}{\bibfnamefont{N.}~\bibnamefont{Yunes}},
  \bibinfo{author}{\bibfnamefont{D.}~\bibnamefont{Psaltis}},
  \bibinfo{author}{\bibfnamefont{F.}~\bibnamefont{Ozel}}, \bibnamefont{and}
  \bibinfo{author}{\bibfnamefont{A.}~\bibnamefont{Loeb}},
  \bibinfo{journal}{Phys. Rev. D} \textbf{\bibinfo{volume}{81}},
  \bibinfo{pages}{064020} (\bibinfo{year}{2010}), \eprint{0912.2736}.

\bibitem[{\citenamefont{Ali-Haimoud and Chen}(2011)}]{Ali-Haimoud:2011zme}
\bibinfo{author}{\bibfnamefont{Y.}~\bibnamefont{Ali-Haimoud}} \bibnamefont{and}
  \bibinfo{author}{\bibfnamefont{Y.}~\bibnamefont{Chen}},
  \bibinfo{journal}{Phys. Rev. D} \textbf{\bibinfo{volume}{84}},
  \bibinfo{pages}{124033} (\bibinfo{year}{2011}), \eprint{1110.5329}.

\bibitem[{\citenamefont{Nakamura et~al.}(2019)\citenamefont{Nakamura, Kikuchi,
  Yamada, Asada, and Yunes}}]{Nakamura:2018yaw}
\bibinfo{author}{\bibfnamefont{Y.}~\bibnamefont{Nakamura}},
  \bibinfo{author}{\bibfnamefont{D.}~\bibnamefont{Kikuchi}},
  \bibinfo{author}{\bibfnamefont{K.}~\bibnamefont{Yamada}},
  \bibinfo{author}{\bibfnamefont{H.}~\bibnamefont{Asada}}, \bibnamefont{and}
  \bibinfo{author}{\bibfnamefont{N.}~\bibnamefont{Yunes}},
  \bibinfo{journal}{Class. Quant. Grav.} \textbf{\bibinfo{volume}{36}},
  \bibinfo{pages}{105006} (\bibinfo{year}{2019}), \eprint{1810.13313}.

\bibitem[{\citenamefont{Everitt et~al.}(2011)}]{Everitt:2011hp}
\bibinfo{author}{\bibfnamefont{C.~W.~F.} \bibnamefont{Everitt}}
  \bibnamefont{et~al.}, \bibinfo{journal}{Phys. Rev. Lett.}
  \textbf{\bibinfo{volume}{106}}, \bibinfo{pages}{221101}
  (\bibinfo{year}{2011}), \eprint{1105.3456}.

\bibitem[{\citenamefont{Silva et~al.}(2021)\citenamefont{Silva, Holgado,
  C\'ardenas-Avenda\~no, and Yunes}}]{Silva:2020acr}
\bibinfo{author}{\bibfnamefont{H.~O.} \bibnamefont{Silva}},
  \bibinfo{author}{\bibfnamefont{A.~M.} \bibnamefont{Holgado}},
  \bibinfo{author}{\bibfnamefont{A.}~\bibnamefont{C\'ardenas-Avenda\~no}},
  \bibnamefont{and} \bibinfo{author}{\bibfnamefont{N.}~\bibnamefont{Yunes}},
  \bibinfo{journal}{Phys. Rev. Lett.} \textbf{\bibinfo{volume}{126}},
  \bibinfo{pages}{181101} (\bibinfo{year}{2021}), \eprint{2004.01253}.

\bibitem[{\citenamefont{Loutrel et~al.}(2019)\citenamefont{Loutrel, Tanaka, and
  Yunes}}]{Loutrel:2018rxs}
\bibinfo{author}{\bibfnamefont{N.}~\bibnamefont{Loutrel}},
  \bibinfo{author}{\bibfnamefont{T.}~\bibnamefont{Tanaka}}, \bibnamefont{and}
  \bibinfo{author}{\bibfnamefont{N.}~\bibnamefont{Yunes}},
  \bibinfo{journal}{Class. Quant. Grav.} \textbf{\bibinfo{volume}{36}},
  \bibinfo{pages}{10LT02} (\bibinfo{year}{2019}), \eprint{1806.07425}.

\bibitem[{\citenamefont{Loutrel and Yunes}(2022)}]{Loutrel:2022tbk}
\bibinfo{author}{\bibfnamefont{N.}~\bibnamefont{Loutrel}} \bibnamefont{and}
  \bibinfo{author}{\bibfnamefont{N.}~\bibnamefont{Yunes}},
  \bibinfo{journal}{Phys. Rev. D} \textbf{\bibinfo{volume}{106}},
  \bibinfo{pages}{064009} (\bibinfo{year}{2022}), \eprint{2205.02675}.

\bibitem[{\citenamefont{Zhang et~al.}(2022)\citenamefont{Zhang, Feng, and
  Gao}}]{Zhang:2022xmm}
\bibinfo{author}{\bibfnamefont{F.}~\bibnamefont{Zhang}},
  \bibinfo{author}{\bibfnamefont{J.-X.} \bibnamefont{Feng}}, \bibnamefont{and}
  \bibinfo{author}{\bibfnamefont{X.}~\bibnamefont{Gao}},
  \bibinfo{journal}{JCAP} \textbf{\bibinfo{volume}{10}}, \bibinfo{pages}{054}
  (\bibinfo{year}{2022}), \eprint{2205.12045}.

\bibitem[{\citenamefont{Perkins et~al.}(2021)\citenamefont{Perkins, Nair,
  Silva, and Yunes}}]{Perkins:2021mhb}
\bibinfo{author}{\bibfnamefont{S.~E.} \bibnamefont{Perkins}},
  \bibinfo{author}{\bibfnamefont{R.}~\bibnamefont{Nair}},
  \bibinfo{author}{\bibfnamefont{H.~O.} \bibnamefont{Silva}}, \bibnamefont{and}
  \bibinfo{author}{\bibfnamefont{N.}~\bibnamefont{Yunes}},
  \bibinfo{journal}{Phys. Rev. D} \textbf{\bibinfo{volume}{104}},
  \bibinfo{pages}{024060} (\bibinfo{year}{2021}), \eprint{2104.11189}.

\bibitem[{\citenamefont{Cardoso and Gualtieri}(2009)}]{Cardoso:2009pk}
\bibinfo{author}{\bibfnamefont{V.}~\bibnamefont{Cardoso}} \bibnamefont{and}
  \bibinfo{author}{\bibfnamefont{L.}~\bibnamefont{Gualtieri}},
  \bibinfo{journal}{Phys. Rev. D} \textbf{\bibinfo{volume}{80}},
  \bibinfo{pages}{064008} (\bibinfo{year}{2009}), \bibinfo{note}{[Erratum:
  Phys.Rev.D 81, 089903 (2010)]}, \eprint{0907.5008}.

\bibitem[{\citenamefont{Konno et~al.}(2009)\citenamefont{Konno, Matsuyama, and
  Tanda}}]{Konno:2009kg}
\bibinfo{author}{\bibfnamefont{K.}~\bibnamefont{Konno}},
  \bibinfo{author}{\bibfnamefont{T.}~\bibnamefont{Matsuyama}},
  \bibnamefont{and} \bibinfo{author}{\bibfnamefont{S.}~\bibnamefont{Tanda}},
  \bibinfo{journal}{Prog. Theor. Phys.} \textbf{\bibinfo{volume}{122}},
  \bibinfo{pages}{561} (\bibinfo{year}{2009}), \eprint{0902.4767}.

\bibitem[{\citenamefont{Yagi et~al.}(2012)\citenamefont{Yagi, Yunes, and
  Tanaka}}]{Yagi:2012ya}
\bibinfo{author}{\bibfnamefont{K.}~\bibnamefont{Yagi}},
  \bibinfo{author}{\bibfnamefont{N.}~\bibnamefont{Yunes}}, \bibnamefont{and}
  \bibinfo{author}{\bibfnamefont{T.}~\bibnamefont{Tanaka}},
  \bibinfo{journal}{Phys. Rev. D} \textbf{\bibinfo{volume}{86}},
  \bibinfo{pages}{044037} (\bibinfo{year}{2012}), \bibinfo{note}{[Erratum:
  Phys.Rev.D 89, 049902 (2014)]}, \eprint{1206.6130}.

\bibitem[{\citenamefont{Maselli et~al.}(2017)\citenamefont{Maselli, Pani,
  Cotesta, Gualtieri, Ferrari, and Stella}}]{Maselli:2017kic}
\bibinfo{author}{\bibfnamefont{A.}~\bibnamefont{Maselli}},
  \bibinfo{author}{\bibfnamefont{P.}~\bibnamefont{Pani}},
  \bibinfo{author}{\bibfnamefont{R.}~\bibnamefont{Cotesta}},
  \bibinfo{author}{\bibfnamefont{L.}~\bibnamefont{Gualtieri}},
  \bibinfo{author}{\bibfnamefont{V.}~\bibnamefont{Ferrari}}, \bibnamefont{and}
  \bibinfo{author}{\bibfnamefont{L.}~\bibnamefont{Stella}},
  \bibinfo{journal}{Astrophys. J.} \textbf{\bibinfo{volume}{843}},
  \bibinfo{pages}{25} (\bibinfo{year}{2017}), \eprint{1703.01472}.

\bibitem[{\citenamefont{Cano and Ruip\'erez}(2019)}]{Cano:2019ore}
\bibinfo{author}{\bibfnamefont{P.~A.} \bibnamefont{Cano}} \bibnamefont{and}
  \bibinfo{author}{\bibfnamefont{A.}~\bibnamefont{Ruip\'erez}},
  \bibinfo{journal}{JHEP} \textbf{\bibinfo{volume}{05}}, \bibinfo{pages}{189}
  (\bibinfo{year}{2019}), \bibinfo{note}{[Erratum: JHEP 03, 187 (2020)]},
  \eprint{1901.01315}.

\bibitem[{\citenamefont{Delsate et~al.}(2018)\citenamefont{Delsate, Herdeiro,
  and Radu}}]{Delsate:2018ome}
\bibinfo{author}{\bibfnamefont{T.}~\bibnamefont{Delsate}},
  \bibinfo{author}{\bibfnamefont{C.}~\bibnamefont{Herdeiro}}, \bibnamefont{and}
  \bibinfo{author}{\bibfnamefont{E.}~\bibnamefont{Radu}},
  \bibinfo{journal}{Phys. Lett. B} \textbf{\bibinfo{volume}{787}},
  \bibinfo{pages}{8} (\bibinfo{year}{2018}), \eprint{1806.06700}.

\bibitem[{\citenamefont{Hartle and Thorne}(1968)}]{Hartle:1968si}
\bibinfo{author}{\bibfnamefont{J.~B.} \bibnamefont{Hartle}} \bibnamefont{and}
  \bibinfo{author}{\bibfnamefont{K.~S.} \bibnamefont{Thorne}},
  \bibinfo{journal}{Astrophys. J.} \textbf{\bibinfo{volume}{153}},
  \bibinfo{pages}{807} (\bibinfo{year}{1968}).

\bibitem[{\citenamefont{Thorne and Hartle}(1984)}]{Thorne:1984mz}
\bibinfo{author}{\bibfnamefont{K.~S.} \bibnamefont{Thorne}} \bibnamefont{and}
  \bibinfo{author}{\bibfnamefont{J.~B.} \bibnamefont{Hartle}},
  \bibinfo{journal}{Phys. Rev. D} \textbf{\bibinfo{volume}{31}},
  \bibinfo{pages}{1815} (\bibinfo{year}{1984}).

\bibitem[{\citenamefont{Alexander et~al.}(2022)\citenamefont{Alexander,
  Gabadadze, Jenks, and Yunes}}]{Alexander:2022avt}
\bibinfo{author}{\bibfnamefont{S.}~\bibnamefont{Alexander}},
  \bibinfo{author}{\bibfnamefont{G.}~\bibnamefont{Gabadadze}},
  \bibinfo{author}{\bibfnamefont{L.}~\bibnamefont{Jenks}}, \bibnamefont{and}
  \bibinfo{author}{\bibfnamefont{N.}~\bibnamefont{Yunes}}
  (\bibinfo{year}{2022}), \eprint{2201.02220}.

\bibitem[{\citenamefont{Harko et~al.}(2010)\citenamefont{Harko, Kovacs, and
  Lobo}}]{Harko:2009kj}
\bibinfo{author}{\bibfnamefont{T.}~\bibnamefont{Harko}},
  \bibinfo{author}{\bibfnamefont{Z.}~\bibnamefont{Kovacs}}, \bibnamefont{and}
  \bibinfo{author}{\bibfnamefont{F.~S.~N.} \bibnamefont{Lobo}},
  \bibinfo{journal}{Class. Quant. Grav.} \textbf{\bibinfo{volume}{27}},
  \bibinfo{pages}{105010} (\bibinfo{year}{2010}), \eprint{0909.1267}.

\bibitem[{\citenamefont{Amarilla et~al.}(2010)\citenamefont{Amarilla, Eiroa,
  and Giribet}}]{Amarilla:2010zq}
\bibinfo{author}{\bibfnamefont{L.}~\bibnamefont{Amarilla}},
  \bibinfo{author}{\bibfnamefont{E.~F.} \bibnamefont{Eiroa}}, \bibnamefont{and}
  \bibinfo{author}{\bibfnamefont{G.}~\bibnamefont{Giribet}},
  \bibinfo{journal}{Phys. Rev. D} \textbf{\bibinfo{volume}{81}},
  \bibinfo{pages}{124045} (\bibinfo{year}{2010}), \eprint{1005.0607}.

\bibitem[{\citenamefont{Chen and Jing}(2010)}]{Chen:2010yx}
\bibinfo{author}{\bibfnamefont{S.}~\bibnamefont{Chen}} \bibnamefont{and}
  \bibinfo{author}{\bibfnamefont{J.}~\bibnamefont{Jing}},
  \bibinfo{journal}{Class. Quant. Grav.} \textbf{\bibinfo{volume}{27}},
  \bibinfo{pages}{225006} (\bibinfo{year}{2010}), \eprint{1005.1325}.

\bibitem[{\citenamefont{Cunha and Herdeiro}(2018)}]{Cunha:2018acu}
\bibinfo{author}{\bibfnamefont{P.~V.~P.} \bibnamefont{Cunha}} \bibnamefont{and}
  \bibinfo{author}{\bibfnamefont{C.~A.~R.} \bibnamefont{Herdeiro}},
  \bibinfo{journal}{Gen. Rel. Grav.} \textbf{\bibinfo{volume}{50}},
  \bibinfo{pages}{42} (\bibinfo{year}{2018}), \eprint{1801.00860}.

\bibitem[{\citenamefont{Perlick and Tsupko}(2022)}]{Perlick:2021aok}
\bibinfo{author}{\bibfnamefont{V.}~\bibnamefont{Perlick}} \bibnamefont{and}
  \bibinfo{author}{\bibfnamefont{O.~Y.} \bibnamefont{Tsupko}},
  \bibinfo{journal}{Phys. Rept.} \textbf{\bibinfo{volume}{947}},
  \bibinfo{pages}{1} (\bibinfo{year}{2022}), \eprint{2105.07101}.

\bibitem[{\citenamefont{Chen et~al.}(2022)\citenamefont{Chen, Jing, Qian, and
  Wang}}]{Chen:2022scf}
\bibinfo{author}{\bibfnamefont{S.}~\bibnamefont{Chen}},
  \bibinfo{author}{\bibfnamefont{J.}~\bibnamefont{Jing}},
  \bibinfo{author}{\bibfnamefont{W.-L.} \bibnamefont{Qian}}, \bibnamefont{and}
  \bibinfo{author}{\bibfnamefont{B.}~\bibnamefont{Wang}},
  \bibinfo{journal}{Sci. China-Phys. Mech. Astron}  (\bibinfo{year}{2022}),
  \eprint{2301.00113}.

\bibitem[{\citenamefont{Cunningham and Bardeen}(1973)}]{cunningham1973optical}
\bibinfo{author}{\bibfnamefont{C.}~\bibnamefont{Cunningham}} \bibnamefont{and}
  \bibinfo{author}{\bibfnamefont{J.~M.} \bibnamefont{Bardeen}},
  \bibinfo{journal}{The Astrophysical Journal} \textbf{\bibinfo{volume}{183}},
  \bibinfo{pages}{237} (\bibinfo{year}{1973}).

\bibitem[{\citenamefont{Abdujabbarov et~al.}(2016)\citenamefont{Abdujabbarov,
  Amir, Ahmedov, and Ghosh}}]{Abdujabbarov:2016hnw}
\bibinfo{author}{\bibfnamefont{A.}~\bibnamefont{Abdujabbarov}},
  \bibinfo{author}{\bibfnamefont{M.}~\bibnamefont{Amir}},
  \bibinfo{author}{\bibfnamefont{B.}~\bibnamefont{Ahmedov}}, \bibnamefont{and}
  \bibinfo{author}{\bibfnamefont{S.~G.} \bibnamefont{Ghosh}},
  \bibinfo{journal}{Phys. Rev. D} \textbf{\bibinfo{volume}{93}},
  \bibinfo{pages}{104004} (\bibinfo{year}{2016}), \eprint{1604.03809}.

\bibitem[{\citenamefont{Grenzebach et~al.}(2014)\citenamefont{Grenzebach,
  Perlick, and L\"ammerzahl}}]{Grenzebach:2014fha}
\bibinfo{author}{\bibfnamefont{A.}~\bibnamefont{Grenzebach}},
  \bibinfo{author}{\bibfnamefont{V.}~\bibnamefont{Perlick}}, \bibnamefont{and}
  \bibinfo{author}{\bibfnamefont{C.}~\bibnamefont{L\"ammerzahl}},
  \bibinfo{journal}{Phys. Rev. D} \textbf{\bibinfo{volume}{89}},
  \bibinfo{pages}{124004} (\bibinfo{year}{2014}), \eprint{1403.5234}.

\bibitem[{\citenamefont{Kumar et~al.}(2019)\citenamefont{Kumar, Ghosh, and
  Wang}}]{Kumar:2019pjp}
\bibinfo{author}{\bibfnamefont{R.}~\bibnamefont{Kumar}},
  \bibinfo{author}{\bibfnamefont{S.~G.} \bibnamefont{Ghosh}}, \bibnamefont{and}
  \bibinfo{author}{\bibfnamefont{A.}~\bibnamefont{Wang}},
  \bibinfo{journal}{Phys. Rev. D} \textbf{\bibinfo{volume}{100}},
  \bibinfo{pages}{124024} (\bibinfo{year}{2019}), \eprint{1912.05154}.

\bibitem[{\citenamefont{Nemmen}(2019)}]{Nemmen:2019idv}
\bibinfo{author}{\bibfnamefont{R.}~\bibnamefont{Nemmen}},
  \bibinfo{journal}{Astrophys. J. Lett.} \textbf{\bibinfo{volume}{880}},
  \bibinfo{pages}{L26} (\bibinfo{year}{2019}), \eprint{1905.02143}.

\bibitem[{\citenamefont{Issaoun et~al.}(2019)}]{Issaoun:2019afg}
\bibinfo{author}{\bibfnamefont{S.}~\bibnamefont{Issaoun}} \bibnamefont{et~al.},
  \bibinfo{journal}{Astrophys. J.} \textbf{\bibinfo{volume}{871}},
  \bibinfo{pages}{30} (\bibinfo{year}{2019}), \eprint{1901.06226}.

\bibitem[{\citenamefont{Srivastava et~al.}(2021)\citenamefont{Srivastava, Chen,
  and Shankaranarayanan}}]{Srivastava:2021imr}
\bibinfo{author}{\bibfnamefont{M.}~\bibnamefont{Srivastava}},
  \bibinfo{author}{\bibfnamefont{Y.}~\bibnamefont{Chen}}, \bibnamefont{and}
  \bibinfo{author}{\bibfnamefont{S.}~\bibnamefont{Shankaranarayanan}},
  \bibinfo{journal}{Phys. Rev. D} \textbf{\bibinfo{volume}{104}},
  \bibinfo{pages}{064034} (\bibinfo{year}{2021}), \eprint{2106.06209}.

\bibitem[{\citenamefont{Wagle et~al.}(2022)\citenamefont{Wagle, Yunes, and
  Silva}}]{Wagle:2021tam}
\bibinfo{author}{\bibfnamefont{P.}~\bibnamefont{Wagle}},
  \bibinfo{author}{\bibfnamefont{N.}~\bibnamefont{Yunes}}, \bibnamefont{and}
  \bibinfo{author}{\bibfnamefont{H.~O.} \bibnamefont{Silva}},
  \bibinfo{journal}{Phys. Rev. D} \textbf{\bibinfo{volume}{105}},
  \bibinfo{pages}{124003} (\bibinfo{year}{2022}), \eprint{2103.09913}.

\bibitem[{\citenamefont{Ghasemi-Nodehi
  et~al.}(2020)\citenamefont{Ghasemi-Nodehi, Azreg-A\"\i{}nou, Jusufi, and
  Jamil}}]{Ghasemi-Nodehi:2020oiz}
\bibinfo{author}{\bibfnamefont{M.}~\bibnamefont{Ghasemi-Nodehi}},
  \bibinfo{author}{\bibfnamefont{M.}~\bibnamefont{Azreg-A\"\i{}nou}},
  \bibinfo{author}{\bibfnamefont{K.}~\bibnamefont{Jusufi}}, \bibnamefont{and}
  \bibinfo{author}{\bibfnamefont{M.}~\bibnamefont{Jamil}},
  \bibinfo{journal}{Phys. Rev. D} \textbf{\bibinfo{volume}{102}},
  \bibinfo{pages}{104032} (\bibinfo{year}{2020}), \eprint{2011.02276}.

\bibitem[{\citenamefont{Gibbons and Werner}(2008)}]{Gibbons:2008rj}
\bibinfo{author}{\bibfnamefont{G.~W.} \bibnamefont{Gibbons}} \bibnamefont{and}
  \bibinfo{author}{\bibfnamefont{M.~C.} \bibnamefont{Werner}},
  \bibinfo{journal}{Class. Quant. Grav.} \textbf{\bibinfo{volume}{25}},
  \bibinfo{pages}{235009} (\bibinfo{year}{2008}), \eprint{0807.0854}.

\bibitem[{\citenamefont{Ishihara et~al.}(2016)\citenamefont{Ishihara, Suzuki,
  Ono, Kitamura, and Asada}}]{Ishihara:2016vdc}
\bibinfo{author}{\bibfnamefont{A.}~\bibnamefont{Ishihara}},
  \bibinfo{author}{\bibfnamefont{Y.}~\bibnamefont{Suzuki}},
  \bibinfo{author}{\bibfnamefont{T.}~\bibnamefont{Ono}},
  \bibinfo{author}{\bibfnamefont{T.}~\bibnamefont{Kitamura}}, \bibnamefont{and}
  \bibinfo{author}{\bibfnamefont{H.}~\bibnamefont{Asada}},
  \bibinfo{journal}{Phys. Rev. D} \textbf{\bibinfo{volume}{94}},
  \bibinfo{pages}{084015} (\bibinfo{year}{2016}), \eprint{1604.08308}.

\bibitem[{\citenamefont{Werner}(2012)}]{Werner:2012rc}
\bibinfo{author}{\bibfnamefont{M.~C.} \bibnamefont{Werner}},
  \bibinfo{journal}{Gen. Rel. Grav.} \textbf{\bibinfo{volume}{44}},
  \bibinfo{pages}{3047} (\bibinfo{year}{2012}), \eprint{1205.3876}.

\bibitem[{\citenamefont{Asada and Kasai}(2000)}]{Asada:2000vn}
\bibinfo{author}{\bibfnamefont{H.}~\bibnamefont{Asada}} \bibnamefont{and}
  \bibinfo{author}{\bibfnamefont{M.}~\bibnamefont{Kasai}},
  \bibinfo{journal}{Prog. Theor. Phys.} \textbf{\bibinfo{volume}{104}},
  \bibinfo{pages}{95} (\bibinfo{year}{2000}), \eprint{astro-ph/0006157}.

\bibitem[{\citenamefont{Ono et~al.}(2017)\citenamefont{Ono, Ishihara, and
  Asada}}]{Ono:2017pie}
\bibinfo{author}{\bibfnamefont{T.}~\bibnamefont{Ono}},
  \bibinfo{author}{\bibfnamefont{A.}~\bibnamefont{Ishihara}}, \bibnamefont{and}
  \bibinfo{author}{\bibfnamefont{H.}~\bibnamefont{Asada}},
  \bibinfo{journal}{Phys. Rev. D} \textbf{\bibinfo{volume}{96}},
  \bibinfo{pages}{104037} (\bibinfo{year}{2017}), \eprint{1704.05615}.

\bibitem[{\citenamefont{Crisnejo
  et~al.}(2019{\natexlab{a}})\citenamefont{Crisnejo, Gallo, and
  Jusufi}}]{Crisnejo:2019ril}
\bibinfo{author}{\bibfnamefont{G.}~\bibnamefont{Crisnejo}},
  \bibinfo{author}{\bibfnamefont{E.}~\bibnamefont{Gallo}}, \bibnamefont{and}
  \bibinfo{author}{\bibfnamefont{K.}~\bibnamefont{Jusufi}},
  \bibinfo{journal}{Phys. Rev. D} \textbf{\bibinfo{volume}{100}},
  \bibinfo{pages}{104045} (\bibinfo{year}{2019}{\natexlab{a}}),
  \eprint{1910.02030}.

\bibitem[{\citenamefont{\"Ovg\"un}(2018)}]{Ovgun:2018fnk}
\bibinfo{author}{\bibfnamefont{A.}~\bibnamefont{\"Ovg\"un}},
  \bibinfo{journal}{Phys. Rev. D} \textbf{\bibinfo{volume}{98}},
  \bibinfo{pages}{044033} (\bibinfo{year}{2018}), \eprint{1805.06296}.

\bibitem[{\citenamefont{Javed et~al.}(2019{\natexlab{a}})\citenamefont{Javed,
  Babar, and \"Ovg\"un}}]{Javed:2019ynm}
\bibinfo{author}{\bibfnamefont{W.}~\bibnamefont{Javed}},
  \bibinfo{author}{\bibfnamefont{R.}~\bibnamefont{Babar}}, \bibnamefont{and}
  \bibinfo{author}{\bibfnamefont{A.}~\bibnamefont{\"Ovg\"un}},
  \bibinfo{journal}{Phys. Rev. D} \textbf{\bibinfo{volume}{100}},
  \bibinfo{pages}{104032} (\bibinfo{year}{2019}{\natexlab{a}}),
  \eprint{1910.11697}.

\bibitem[{\citenamefont{Javed et~al.}(2019{\natexlab{b}})\citenamefont{Javed,
  Abbas, and \"Ovg\"un}}]{Javed:2019kon}
\bibinfo{author}{\bibfnamefont{W.}~\bibnamefont{Javed}},
  \bibinfo{author}{\bibfnamefont{J.}~\bibnamefont{Abbas}}, \bibnamefont{and}
  \bibinfo{author}{\bibfnamefont{A.}~\bibnamefont{\"Ovg\"un}},
  \bibinfo{journal}{Eur. Phys. J. C} \textbf{\bibinfo{volume}{79}},
  \bibinfo{pages}{694} (\bibinfo{year}{2019}{\natexlab{b}}),
  \eprint{1908.09632}.

\bibitem[{\citenamefont{Crisnejo
  et~al.}(2019{\natexlab{b}})\citenamefont{Crisnejo, Gallo, and
  Villanueva}}]{Crisnejo:2019xtp}
\bibinfo{author}{\bibfnamefont{G.}~\bibnamefont{Crisnejo}},
  \bibinfo{author}{\bibfnamefont{E.}~\bibnamefont{Gallo}}, \bibnamefont{and}
  \bibinfo{author}{\bibfnamefont{J.~R.} \bibnamefont{Villanueva}},
  \bibinfo{journal}{Phys. Rev. D} \textbf{\bibinfo{volume}{100}},
  \bibinfo{pages}{044006} (\bibinfo{year}{2019}{\natexlab{b}}),
  \eprint{1905.02125}.

\bibitem[{\citenamefont{Crisnejo
  et~al.}(2019{\natexlab{c}})\citenamefont{Crisnejo, Gallo, and
  Rogers}}]{Crisnejo:2018ppm}
\bibinfo{author}{\bibfnamefont{G.}~\bibnamefont{Crisnejo}},
  \bibinfo{author}{\bibfnamefont{E.}~\bibnamefont{Gallo}}, \bibnamefont{and}
  \bibinfo{author}{\bibfnamefont{A.}~\bibnamefont{Rogers}},
  \bibinfo{journal}{Phys. Rev. D} \textbf{\bibinfo{volume}{99}},
  \bibinfo{pages}{124001} (\bibinfo{year}{2019}{\natexlab{c}}),
  \eprint{1807.00724}.

\bibitem[{\citenamefont{Zhu et~al.}(2019)\citenamefont{Zhu, Wu, Jamil, and
  Jusufi}}]{Zhu:2019ura}
\bibinfo{author}{\bibfnamefont{T.}~\bibnamefont{Zhu}},
  \bibinfo{author}{\bibfnamefont{Q.}~\bibnamefont{Wu}},
  \bibinfo{author}{\bibfnamefont{M.}~\bibnamefont{Jamil}}, \bibnamefont{and}
  \bibinfo{author}{\bibfnamefont{K.}~\bibnamefont{Jusufi}},
  \bibinfo{journal}{Phys. Rev. D} \textbf{\bibinfo{volume}{100}},
  \bibinfo{pages}{044055} (\bibinfo{year}{2019}), \eprint{1906.05673}.

\bibitem[{\citenamefont{Qiao and Zhou}(2021)}]{Qiao:2021trw}
\bibinfo{author}{\bibfnamefont{C.-K.} \bibnamefont{Qiao}} \bibnamefont{and}
  \bibinfo{author}{\bibfnamefont{M.}~\bibnamefont{Zhou}}
  (\bibinfo{year}{2021}), \eprint{2109.05828}.

\bibitem[{\citenamefont{Ono and Asada}(2019)}]{Ono:2019hkw}
\bibinfo{author}{\bibfnamefont{T.}~\bibnamefont{Ono}} \bibnamefont{and}
  \bibinfo{author}{\bibfnamefont{H.}~\bibnamefont{Asada}},
  \bibinfo{journal}{Universe} \textbf{\bibinfo{volume}{5}},
  \bibinfo{pages}{218} (\bibinfo{year}{2019}), \eprint{1906.02414}.

\bibitem[{\citenamefont{Kumar et~al.}(2020{\natexlab{b}})\citenamefont{Kumar,
  Singh, and Ghosh}}]{Kumar:2019ohr}
\bibinfo{author}{\bibfnamefont{R.}~\bibnamefont{Kumar}},
  \bibinfo{author}{\bibfnamefont{B.~P.} \bibnamefont{Singh}}, \bibnamefont{and}
  \bibinfo{author}{\bibfnamefont{S.~G.} \bibnamefont{Ghosh}},
  \bibinfo{journal}{Annals Phys.} \textbf{\bibinfo{volume}{420}},
  \bibinfo{pages}{168252} (\bibinfo{year}{2020}{\natexlab{b}}),
  \eprint{1904.07652}.

\bibitem[{\citenamefont{Carter}(1968{\natexlab{a}})}]{Carter:1968rr}
\bibinfo{author}{\bibfnamefont{B.}~\bibnamefont{Carter}},
  \bibinfo{journal}{Phys. Rev.} \textbf{\bibinfo{volume}{174}},
  \bibinfo{pages}{1559} (\bibinfo{year}{1968}{\natexlab{a}}).

\bibitem[{\citenamefont{Carter}(1968{\natexlab{b}})}]{Carter:1968ks}
\bibinfo{author}{\bibfnamefont{B.}~\bibnamefont{Carter}},
  \bibinfo{journal}{Commun. Math. Phys.} \textbf{\bibinfo{volume}{10}},
  \bibinfo{pages}{280} (\bibinfo{year}{1968}{\natexlab{b}}).

\bibitem[{\citenamefont{Chandrasekhar}(1984)}]{Chandrasekhar:1984siy}
\bibinfo{author}{\bibfnamefont{S.}~\bibnamefont{Chandrasekhar}},
  \bibinfo{journal}{Fundam. Theor. Phys.} \textbf{\bibinfo{volume}{9}},
  \bibinfo{pages}{5} (\bibinfo{year}{1984}).

\end{thebibliography}


\begin{thebibliography}{47}
\expandafter\ifx\csname natexlab\endcsname\relax\def\natexlab#1{#1}\fi
\expandafter\ifx\csname bibnamefont\endcsname\relax
  \def\bibnamefont#1{#1}\fi
\expandafter\ifx\csname bibfnamefont\endcsname\relax
  \def\bibfnamefont#1{#1}\fi
\expandafter\ifx\csname citenamefont\endcsname\relax
  \def\citenamefont#1{#1}\fi
\expandafter\ifx\csname url\endcsname\relax
  \def\url#1{\texttt{#1}}\fi
\expandafter\ifx\csname urlprefix\endcsname\relax\def\urlprefix{URL }\fi
\providecommand{\bibinfo}[2]{#2}
\providecommand{\eprint}[2][]{\url{#2}}

\bibitem[{\citenamefont{Abbott et~al.}(2016)}]{LIGOScientific:2016aoc}
\bibinfo{author}{\bibfnamefont{B.~P.} \bibnamefont{Abbott}}
  \bibnamefont{et~al.} (\bibinfo{collaboration}{LIGO Scientific, Virgo}),
  \bibinfo{journal}{Phys. Rev. Lett.} \textbf{\bibinfo{volume}{116}},
  \bibinfo{pages}{061102} (\bibinfo{year}{2016}), \eprint{1602.03837}.

\bibitem[{\citenamefont{Akiyama et~al.}(2019)}]{EventHorizonTelescope:2019dse}
\bibinfo{author}{\bibfnamefont{K.}~\bibnamefont{Akiyama}} \bibnamefont{et~al.}
  (\bibinfo{collaboration}{Event Horizon Telescope}),
  \bibinfo{journal}{Astrophys. J. Lett.} \textbf{\bibinfo{volume}{875}},
  \bibinfo{pages}{L1} (\bibinfo{year}{2019}), \eprint{1906.11238}.

\bibitem[{\citenamefont{Akiyama et~al.}(2022)}]{EventHorizonTelescope:2022xnr}
\bibinfo{author}{\bibfnamefont{K.}~\bibnamefont{Akiyama}} \bibnamefont{et~al.}
  (\bibinfo{collaboration}{Event Horizon Telescope}),
  \bibinfo{journal}{Astrophys. J. Lett.} \textbf{\bibinfo{volume}{930}},
  \bibinfo{pages}{L12} (\bibinfo{year}{2022}).

\bibitem[{\citenamefont{Synge}(1966)}]{Synge:1966okc}
\bibinfo{author}{\bibfnamefont{J.~L.} \bibnamefont{Synge}},
  \bibinfo{journal}{Mon. Not. Roy. Astron. Soc.}
  \textbf{\bibinfo{volume}{131}}, \bibinfo{pages}{463} (\bibinfo{year}{1966}).

\bibitem[{\citenamefont{Luminet}(1979)}]{Luminet:1979nyg}
\bibinfo{author}{\bibfnamefont{J.~P.} \bibnamefont{Luminet}},
  \bibinfo{journal}{Astron. Astrophys.} \textbf{\bibinfo{volume}{75}},
  \bibinfo{pages}{228} (\bibinfo{year}{1979}).

\bibitem[{\citenamefont{Bardeen}(1973)}]{bardeen1973houches}
\bibinfo{author}{\bibfnamefont{J.}~\bibnamefont{Bardeen}},
  \emph{\bibinfo{title}{Les houches summer school of theoretical physics: Black
  holes}} (\bibinfo{year}{1973}).

\bibitem[{\citenamefont{Cunha and Herdeiro}(2018)}]{Cunha:2018acu}
\bibinfo{author}{\bibfnamefont{P.~V.~P.} \bibnamefont{Cunha}} \bibnamefont{and}
  \bibinfo{author}{\bibfnamefont{C.~A.~R.} \bibnamefont{Herdeiro}},
  \bibinfo{journal}{Gen. Rel. Grav.} \textbf{\bibinfo{volume}{50}},
  \bibinfo{pages}{42} (\bibinfo{year}{2018}), \eprint{1801.00860}.

\bibitem[{\citenamefont{Perlick and Tsupko}(2022)}]{Perlick:2021aok}
\bibinfo{author}{\bibfnamefont{V.}~\bibnamefont{Perlick}} \bibnamefont{and}
  \bibinfo{author}{\bibfnamefont{O.~Y.} \bibnamefont{Tsupko}},
  \bibinfo{journal}{Phys. Rept.} \textbf{\bibinfo{volume}{947}},
  \bibinfo{pages}{1} (\bibinfo{year}{2022}), \eprint{2105.07101}.

\bibitem[{\citenamefont{Chen et~al.}(2022)\citenamefont{Chen, Jing, Qian, and
  Wang}}]{Chen:2022scf}
\bibinfo{author}{\bibfnamefont{S.}~\bibnamefont{Chen}},
  \bibinfo{author}{\bibfnamefont{J.}~\bibnamefont{Jing}},
  \bibinfo{author}{\bibfnamefont{W.-L.} \bibnamefont{Qian}}, \bibnamefont{and}
  \bibinfo{author}{\bibfnamefont{B.}~\bibnamefont{Wang}},
  \bibinfo{journal}{Sci. China-Phys. Mech. Astron}  (\bibinfo{year}{2022}),
  \eprint{2301.00113}.

\bibitem[{\citenamefont{Hioki and Maeda}(2009)}]{Hioki:2009na}
\bibinfo{author}{\bibfnamefont{K.}~\bibnamefont{Hioki}} \bibnamefont{and}
  \bibinfo{author}{\bibfnamefont{K.-i.} \bibnamefont{Maeda}},
  \bibinfo{journal}{Phys. Rev. D} \textbf{\bibinfo{volume}{80}},
  \bibinfo{pages}{024042} (\bibinfo{year}{2009}), \eprint{0904.3575}.

\bibitem[{\citenamefont{Kramer et~al.}(2004)\citenamefont{Kramer, Backer,
  Cordes, Lazio, Stappers, and Johnston}}]{Kramer:2004hd}
\bibinfo{author}{\bibfnamefont{M.}~\bibnamefont{Kramer}},
  \bibinfo{author}{\bibfnamefont{D.~C.} \bibnamefont{Backer}},
  \bibinfo{author}{\bibfnamefont{J.~M.} \bibnamefont{Cordes}},
  \bibinfo{author}{\bibfnamefont{T.~J.~W.} \bibnamefont{Lazio}},
  \bibinfo{author}{\bibfnamefont{B.~W.} \bibnamefont{Stappers}},
  \bibnamefont{and} \bibinfo{author}{\bibfnamefont{S.}~\bibnamefont{Johnston}},
  \bibinfo{journal}{New Astron. Rev.} \textbf{\bibinfo{volume}{48}},
  \bibinfo{pages}{993} (\bibinfo{year}{2004}), \eprint{astro-ph/0409379}.

\bibitem[{\citenamefont{Wei and Liu}(2013)}]{Wei:2013kza}
\bibinfo{author}{\bibfnamefont{S.-W.} \bibnamefont{Wei}} \bibnamefont{and}
  \bibinfo{author}{\bibfnamefont{Y.-X.} \bibnamefont{Liu}},
  \bibinfo{journal}{JCAP} \textbf{\bibinfo{volume}{11}}, \bibinfo{pages}{063}
  (\bibinfo{year}{2013}), \eprint{1311.4251}.

\bibitem[{\citenamefont{Tsupko}(2017)}]{Tsupko:2017rdo}
\bibinfo{author}{\bibfnamefont{O.~Y.} \bibnamefont{Tsupko}},
  \bibinfo{journal}{Phys. Rev. D} \textbf{\bibinfo{volume}{95}},
  \bibinfo{pages}{104058} (\bibinfo{year}{2017}), \eprint{1702.04005}.

\bibitem[{\citenamefont{Allahyari et~al.}(2020)\citenamefont{Allahyari,
  Khodadi, Vagnozzi, and Mota}}]{Allahyari:2019jqz}
\bibinfo{author}{\bibfnamefont{A.}~\bibnamefont{Allahyari}},
  \bibinfo{author}{\bibfnamefont{M.}~\bibnamefont{Khodadi}},
  \bibinfo{author}{\bibfnamefont{S.}~\bibnamefont{Vagnozzi}}, \bibnamefont{and}
  \bibinfo{author}{\bibfnamefont{D.~F.} \bibnamefont{Mota}},
  \bibinfo{journal}{JCAP} \textbf{\bibinfo{volume}{02}}, \bibinfo{pages}{003}
  (\bibinfo{year}{2020}), \eprint{1912.08231}.

\bibitem[{\citenamefont{Hou et~al.}(2021)\citenamefont{Hou, Guo, and
  Chen}}]{Hou:2021okc}
\bibinfo{author}{\bibfnamefont{Y.}~\bibnamefont{Hou}},
  \bibinfo{author}{\bibfnamefont{M.}~\bibnamefont{Guo}}, \bibnamefont{and}
  \bibinfo{author}{\bibfnamefont{B.}~\bibnamefont{Chen}},
  \bibinfo{journal}{Phys. Rev. D} \textbf{\bibinfo{volume}{104}},
  \bibinfo{pages}{024001} (\bibinfo{year}{2021}), \eprint{2103.04369}.

\bibitem[{\citenamefont{Gan et~al.}(2021)\citenamefont{Gan, Wang, Wu, and
  Yang}}]{Gan:2021pwu}
\bibinfo{author}{\bibfnamefont{Q.}~\bibnamefont{Gan}},
  \bibinfo{author}{\bibfnamefont{P.}~\bibnamefont{Wang}},
  \bibinfo{author}{\bibfnamefont{H.}~\bibnamefont{Wu}}, \bibnamefont{and}
  \bibinfo{author}{\bibfnamefont{H.}~\bibnamefont{Yang}},
  \bibinfo{journal}{Phys. Rev. D} \textbf{\bibinfo{volume}{104}},
  \bibinfo{pages}{024003} (\bibinfo{year}{2021}), \eprint{2104.08703}.

\bibitem[{\citenamefont{Khodadi et~al.}(2020)\citenamefont{Khodadi, Allahyari,
  Vagnozzi, and Mota}}]{Khodadi:2020jij}
\bibinfo{author}{\bibfnamefont{M.}~\bibnamefont{Khodadi}},
  \bibinfo{author}{\bibfnamefont{A.}~\bibnamefont{Allahyari}},
  \bibinfo{author}{\bibfnamefont{S.}~\bibnamefont{Vagnozzi}}, \bibnamefont{and}
  \bibinfo{author}{\bibfnamefont{D.~F.} \bibnamefont{Mota}},
  \bibinfo{journal}{JCAP} \textbf{\bibinfo{volume}{09}}, \bibinfo{pages}{026}
  (\bibinfo{year}{2020}), \eprint{2005.05992}.

\bibitem[{\citenamefont{Bad\'\i{}a and Eiroa}(2021)}]{Badia:2021kpk}
\bibinfo{author}{\bibfnamefont{J.}~\bibnamefont{Bad\'\i{}a}} \bibnamefont{and}
  \bibinfo{author}{\bibfnamefont{E.~F.} \bibnamefont{Eiroa}},
  \bibinfo{journal}{Phys. Rev. D} \textbf{\bibinfo{volume}{104}},
  \bibinfo{pages}{084055} (\bibinfo{year}{2021}), \eprint{2106.07601}.

\bibitem[{\citenamefont{Meng et~al.}(2022)\citenamefont{Meng, Kuang, and
  Tang}}]{Meng:2022kjs}
\bibinfo{author}{\bibfnamefont{Y.}~\bibnamefont{Meng}},
  \bibinfo{author}{\bibfnamefont{X.-M.} \bibnamefont{Kuang}}, \bibnamefont{and}
  \bibinfo{author}{\bibfnamefont{Z.-Y.} \bibnamefont{Tang}},
  \bibinfo{journal}{Phys. Rev. D} \textbf{\bibinfo{volume}{106}},
  \bibinfo{pages}{064006} (\bibinfo{year}{2022}), \eprint{2204.00897}.

\bibitem[{\citenamefont{Kuang and \"Ovg\"un}(2022)}]{Kuang:2022xjp}
\bibinfo{author}{\bibfnamefont{X.-M.} \bibnamefont{Kuang}} \bibnamefont{and}
  \bibinfo{author}{\bibfnamefont{A.}~\bibnamefont{\"Ovg\"un}},
  \bibinfo{journal}{Annals Phys.} \textbf{\bibinfo{volume}{447}},
  \bibinfo{pages}{169147} (\bibinfo{year}{2022}), \eprint{2205.11003}.

\bibitem[{\citenamefont{Cardoso et~al.}(2009)\citenamefont{Cardoso, Miranda,
  Berti, Witek, and Zanchin}}]{Cardoso:2008bp}
\bibinfo{author}{\bibfnamefont{V.}~\bibnamefont{Cardoso}},
  \bibinfo{author}{\bibfnamefont{A.~S.} \bibnamefont{Miranda}},
  \bibinfo{author}{\bibfnamefont{E.}~\bibnamefont{Berti}},
  \bibinfo{author}{\bibfnamefont{H.}~\bibnamefont{Witek}}, \bibnamefont{and}
  \bibinfo{author}{\bibfnamefont{V.~T.} \bibnamefont{Zanchin}},
  \bibinfo{journal}{Phys. Rev. D} \textbf{\bibinfo{volume}{79}},
  \bibinfo{pages}{064016} (\bibinfo{year}{2009}), \eprint{0812.1806}.

\bibitem[{\citenamefont{Jusufi}(2020{\natexlab{a}})}]{Jusufi:2019ltj}
\bibinfo{author}{\bibfnamefont{K.}~\bibnamefont{Jusufi}},
  \bibinfo{journal}{Phys. Rev. D} \textbf{\bibinfo{volume}{101}},
  \bibinfo{pages}{084055} (\bibinfo{year}{2020}{\natexlab{a}}),
  \eprint{1912.13320}.

\bibitem[{\citenamefont{Liu et~al.}(2020)\citenamefont{Liu, Zhu, Wu, Jusufi,
  Jamil, Azreg-A\"\i{}nou, and Wang}}]{Liu:2020ola}
\bibinfo{author}{\bibfnamefont{C.}~\bibnamefont{Liu}},
  \bibinfo{author}{\bibfnamefont{T.}~\bibnamefont{Zhu}},
  \bibinfo{author}{\bibfnamefont{Q.}~\bibnamefont{Wu}},
  \bibinfo{author}{\bibfnamefont{K.}~\bibnamefont{Jusufi}},
  \bibinfo{author}{\bibfnamefont{M.}~\bibnamefont{Jamil}},
  \bibinfo{author}{\bibfnamefont{M.}~\bibnamefont{Azreg-A\"\i{}nou}},
  \bibnamefont{and} \bibinfo{author}{\bibfnamefont{A.}~\bibnamefont{Wang}},
  \bibinfo{journal}{Phys. Rev. D} \textbf{\bibinfo{volume}{101}},
  \bibinfo{pages}{084001} (\bibinfo{year}{2020}), \bibinfo{note}{[Erratum:
  Phys.Rev.D 103, 089902 (2021)]}, \eprint{2003.00477}.

\bibitem[{\citenamefont{Jusufi}(2020{\natexlab{b}})}]{Jusufi:2020dhz}
\bibinfo{author}{\bibfnamefont{K.}~\bibnamefont{Jusufi}},
  \bibinfo{journal}{Phys. Rev. D} \textbf{\bibinfo{volume}{101}},
  \bibinfo{pages}{124063} (\bibinfo{year}{2020}{\natexlab{b}}),
  \eprint{2004.04664}.

\bibitem[{\citenamefont{Alexander and Yunes}(2009)}]{Alexander:2009tp}
\bibinfo{author}{\bibfnamefont{S.}~\bibnamefont{Alexander}} \bibnamefont{and}
  \bibinfo{author}{\bibfnamefont{N.}~\bibnamefont{Yunes}},
  \bibinfo{journal}{Phys. Rept.} \textbf{\bibinfo{volume}{480}},
  \bibinfo{pages}{1} (\bibinfo{year}{2009}), \eprint{0907.2562}.

\bibitem[{\citenamefont{Alexander et~al.}(2008)\citenamefont{Alexander, Finn,
  and Yunes}}]{Alexander:2007kv}
\bibinfo{author}{\bibfnamefont{S.}~\bibnamefont{Alexander}},
  \bibinfo{author}{\bibfnamefont{L.~S.} \bibnamefont{Finn}}, \bibnamefont{and}
  \bibinfo{author}{\bibfnamefont{N.}~\bibnamefont{Yunes}},
  \bibinfo{journal}{Phys. Rev. D} \textbf{\bibinfo{volume}{78}},
  \bibinfo{pages}{066005} (\bibinfo{year}{2008}), \eprint{0712.2542}.

\bibitem[{\citenamefont{Yunes and Pretorius}(2009)}]{Yunes:2009hc}
\bibinfo{author}{\bibfnamefont{N.}~\bibnamefont{Yunes}} \bibnamefont{and}
  \bibinfo{author}{\bibfnamefont{F.}~\bibnamefont{Pretorius}},
  \bibinfo{journal}{Phys. Rev. D} \textbf{\bibinfo{volume}{79}},
  \bibinfo{pages}{084043} (\bibinfo{year}{2009}), \eprint{0902.4669}.

\bibitem[{\citenamefont{Cardoso and Gualtieri}(2009)}]{Cardoso:2009pk}
\bibinfo{author}{\bibfnamefont{V.}~\bibnamefont{Cardoso}} \bibnamefont{and}
  \bibinfo{author}{\bibfnamefont{L.}~\bibnamefont{Gualtieri}},
  \bibinfo{journal}{Phys. Rev. D} \textbf{\bibinfo{volume}{80}},
  \bibinfo{pages}{064008} (\bibinfo{year}{2009}), \bibinfo{note}{[Erratum:
  Phys.Rev.D 81, 089903 (2010)]}, \eprint{0907.5008}.

\bibitem[{\citenamefont{Konno et~al.}(2009)\citenamefont{Konno, Matsuyama, and
  Tanda}}]{Konno:2009kg}
\bibinfo{author}{\bibfnamefont{K.}~\bibnamefont{Konno}},
  \bibinfo{author}{\bibfnamefont{T.}~\bibnamefont{Matsuyama}},
  \bibnamefont{and} \bibinfo{author}{\bibfnamefont{S.}~\bibnamefont{Tanda}},
  \bibinfo{journal}{Prog. Theor. Phys.} \textbf{\bibinfo{volume}{122}},
  \bibinfo{pages}{561} (\bibinfo{year}{2009}), \eprint{0902.4767}.

\bibitem[{\citenamefont{Cano and Ruip\'erez}(2019)}]{Cano:2019ore}
\bibinfo{author}{\bibfnamefont{P.~A.} \bibnamefont{Cano}} \bibnamefont{and}
  \bibinfo{author}{\bibfnamefont{A.}~\bibnamefont{Ruip\'erez}},
  \bibinfo{journal}{JHEP} \textbf{\bibinfo{volume}{05}}, \bibinfo{pages}{189}
  (\bibinfo{year}{2019}), \bibinfo{note}{[Erratum: JHEP 03, 187 (2020)]},
  \eprint{1901.01315}.

\bibitem[{\citenamefont{Delsate et~al.}(2018)\citenamefont{Delsate, Herdeiro,
  and Radu}}]{Delsate:2018ome}
\bibinfo{author}{\bibfnamefont{T.}~\bibnamefont{Delsate}},
  \bibinfo{author}{\bibfnamefont{C.}~\bibnamefont{Herdeiro}}, \bibnamefont{and}
  \bibinfo{author}{\bibfnamefont{E.}~\bibnamefont{Radu}},
  \bibinfo{journal}{Phys. Lett. B} \textbf{\bibinfo{volume}{787}},
  \bibinfo{pages}{8} (\bibinfo{year}{2018}), \eprint{1806.06700}.

\bibitem[{\citenamefont{Alexander et~al.}(2022)\citenamefont{Alexander,
  Gabadadze, Jenks, and Yunes}}]{Alexander:2022avt}
\bibinfo{author}{\bibfnamefont{S.}~\bibnamefont{Alexander}},
  \bibinfo{author}{\bibfnamefont{G.}~\bibnamefont{Gabadadze}},
  \bibinfo{author}{\bibfnamefont{L.}~\bibnamefont{Jenks}}, \bibnamefont{and}
  \bibinfo{author}{\bibfnamefont{N.}~\bibnamefont{Yunes}}
  (\bibinfo{year}{2022}), \eprint{2201.02220}.

\bibitem[{\citenamefont{Harko et~al.}(2010)\citenamefont{Harko, Kovacs, and
  Lobo}}]{Harko:2009kj}
\bibinfo{author}{\bibfnamefont{T.}~\bibnamefont{Harko}},
  \bibinfo{author}{\bibfnamefont{Z.}~\bibnamefont{Kovacs}}, \bibnamefont{and}
  \bibinfo{author}{\bibfnamefont{F.~S.~N.} \bibnamefont{Lobo}},
  \bibinfo{journal}{Class. Quant. Grav.} \textbf{\bibinfo{volume}{27}},
  \bibinfo{pages}{105010} (\bibinfo{year}{2010}), \eprint{0909.1267}.

\bibitem[{\citenamefont{Amarilla et~al.}(2010)\citenamefont{Amarilla, Eiroa,
  and Giribet}}]{Amarilla:2010zq}
\bibinfo{author}{\bibfnamefont{L.}~\bibnamefont{Amarilla}},
  \bibinfo{author}{\bibfnamefont{E.~F.} \bibnamefont{Eiroa}}, \bibnamefont{and}
  \bibinfo{author}{\bibfnamefont{G.}~\bibnamefont{Giribet}},
  \bibinfo{journal}{Phys. Rev. D} \textbf{\bibinfo{volume}{81}},
  \bibinfo{pages}{124045} (\bibinfo{year}{2010}), \eprint{1005.0607}.

\bibitem[{\citenamefont{Chen and Jing}(2010)}]{Chen:2010yx}
\bibinfo{author}{\bibfnamefont{S.}~\bibnamefont{Chen}} \bibnamefont{and}
  \bibinfo{author}{\bibfnamefont{J.}~\bibnamefont{Jing}},
  \bibinfo{journal}{Class. Quant. Grav.} \textbf{\bibinfo{volume}{27}},
  \bibinfo{pages}{225006} (\bibinfo{year}{2010}), \eprint{1005.1325}.

\bibitem[{\citenamefont{Abdujabbarov et~al.}(2016)\citenamefont{Abdujabbarov,
  Amir, Ahmedov, and Ghosh}}]{Abdujabbarov:2016hnw}
\bibinfo{author}{\bibfnamefont{A.}~\bibnamefont{Abdujabbarov}},
  \bibinfo{author}{\bibfnamefont{M.}~\bibnamefont{Amir}},
  \bibinfo{author}{\bibfnamefont{B.}~\bibnamefont{Ahmedov}}, \bibnamefont{and}
  \bibinfo{author}{\bibfnamefont{S.~G.} \bibnamefont{Ghosh}},
  \bibinfo{journal}{Phys. Rev. D} \textbf{\bibinfo{volume}{93}},
  \bibinfo{pages}{104004} (\bibinfo{year}{2016}), \eprint{1604.03809}.

\bibitem[{\citenamefont{Kumar and Ghosh}(2020{\natexlab{a}})}]{Kumar:2018ple}
\bibinfo{author}{\bibfnamefont{R.}~\bibnamefont{Kumar}} \bibnamefont{and}
  \bibinfo{author}{\bibfnamefont{S.~G.} \bibnamefont{Ghosh}},
  \bibinfo{journal}{Astrophys. J.} \textbf{\bibinfo{volume}{892}},
  \bibinfo{pages}{78} (\bibinfo{year}{2020}{\natexlab{a}}),
  \eprint{1811.01260}.

\bibitem[{\citenamefont{Nemmen}(2019)}]{Nemmen:2019idv}
\bibinfo{author}{\bibfnamefont{R.}~\bibnamefont{Nemmen}},
  \bibinfo{journal}{Astrophys. J. Lett.} \textbf{\bibinfo{volume}{880}},
  \bibinfo{pages}{L26} (\bibinfo{year}{2019}), \eprint{1905.02143}.

\bibitem[{\citenamefont{Issaoun et~al.}(2019)}]{Issaoun:2019afg}
\bibinfo{author}{\bibfnamefont{S.}~\bibnamefont{Issaoun}} \bibnamefont{et~al.},
  \bibinfo{journal}{Astrophys. J.} \textbf{\bibinfo{volume}{871}},
  \bibinfo{pages}{30} (\bibinfo{year}{2019}), \eprint{1901.06226}.

\bibitem[{\citenamefont{Kumar and Ghosh}(2020{\natexlab{b}})}]{Kumar:2020owy}
\bibinfo{author}{\bibfnamefont{R.}~\bibnamefont{Kumar}} \bibnamefont{and}
  \bibinfo{author}{\bibfnamefont{S.~G.} \bibnamefont{Ghosh}},
  \bibinfo{journal}{JCAP} \textbf{\bibinfo{volume}{07}}, \bibinfo{pages}{053}
  (\bibinfo{year}{2020}{\natexlab{b}}), \eprint{2003.08927}.

\bibitem[{\citenamefont{Srivastava et~al.}(2021)\citenamefont{Srivastava, Chen,
  and Shankaranarayanan}}]{Srivastava:2021imr}
\bibinfo{author}{\bibfnamefont{M.}~\bibnamefont{Srivastava}},
  \bibinfo{author}{\bibfnamefont{Y.}~\bibnamefont{Chen}}, \bibnamefont{and}
  \bibinfo{author}{\bibfnamefont{S.}~\bibnamefont{Shankaranarayanan}},
  \bibinfo{journal}{Phys. Rev. D} \textbf{\bibinfo{volume}{104}},
  \bibinfo{pages}{064034} (\bibinfo{year}{2021}), \eprint{2106.06209}.

\bibitem[{\citenamefont{Wagle et~al.}(2022)\citenamefont{Wagle, Yunes, and
  Silva}}]{Wagle:2021tam}
\bibinfo{author}{\bibfnamefont{P.}~\bibnamefont{Wagle}},
  \bibinfo{author}{\bibfnamefont{N.}~\bibnamefont{Yunes}}, \bibnamefont{and}
  \bibinfo{author}{\bibfnamefont{H.~O.} \bibnamefont{Silva}},
  \bibinfo{journal}{Phys. Rev. D} \textbf{\bibinfo{volume}{105}},
  \bibinfo{pages}{124003} (\bibinfo{year}{2022}), \eprint{2103.09913}.

\bibitem[{\citenamefont{Ghasemi-Nodehi
  et~al.}(2020)\citenamefont{Ghasemi-Nodehi, Azreg-A\"\i{}nou, Jusufi, and
  Jamil}}]{Ghasemi-Nodehi:2020oiz}
\bibinfo{author}{\bibfnamefont{M.}~\bibnamefont{Ghasemi-Nodehi}},
  \bibinfo{author}{\bibfnamefont{M.}~\bibnamefont{Azreg-A\"\i{}nou}},
  \bibinfo{author}{\bibfnamefont{K.}~\bibnamefont{Jusufi}}, \bibnamefont{and}
  \bibinfo{author}{\bibfnamefont{M.}~\bibnamefont{Jamil}},
  \bibinfo{journal}{Phys. Rev. D} \textbf{\bibinfo{volume}{102}},
  \bibinfo{pages}{104032} (\bibinfo{year}{2020}), \eprint{2011.02276}.

\bibitem[{\citenamefont{Gibbons and Werner}(2008)}]{Gibbons:2008rj}
\bibinfo{author}{\bibfnamefont{G.~W.} \bibnamefont{Gibbons}} \bibnamefont{and}
  \bibinfo{author}{\bibfnamefont{M.~C.} \bibnamefont{Werner}},
  \bibinfo{journal}{Class. Quant. Grav.} \textbf{\bibinfo{volume}{25}},
  \bibinfo{pages}{235009} (\bibinfo{year}{2008}), \eprint{0807.0854}.

\bibitem[{\citenamefont{Ishihara et~al.}(2016)\citenamefont{Ishihara, Suzuki,
  Ono, Kitamura, and Asada}}]{Ishihara:2016vdc}
\bibinfo{author}{\bibfnamefont{A.}~\bibnamefont{Ishihara}},
  \bibinfo{author}{\bibfnamefont{Y.}~\bibnamefont{Suzuki}},
  \bibinfo{author}{\bibfnamefont{T.}~\bibnamefont{Ono}},
  \bibinfo{author}{\bibfnamefont{T.}~\bibnamefont{Kitamura}}, \bibnamefont{and}
  \bibinfo{author}{\bibfnamefont{H.}~\bibnamefont{Asada}},
  \bibinfo{journal}{Phys. Rev. D} \textbf{\bibinfo{volume}{94}},
  \bibinfo{pages}{084015} (\bibinfo{year}{2016}), \eprint{1604.08308}.

\bibitem[{\citenamefont{Werner}(2012)}]{Werner:2012rc}
\bibinfo{author}{\bibfnamefont{M.~C.} \bibnamefont{Werner}},
  \bibinfo{journal}{Gen. Rel. Grav.} \textbf{\bibinfo{volume}{44}},
  \bibinfo{pages}{3047} (\bibinfo{year}{2012}), \eprint{1205.3876}.

\bibitem[{\citenamefont{Ono and Asada}(2019)}]{Ono:2019hkw}
\bibinfo{author}{\bibfnamefont{T.}~\bibnamefont{Ono}} \bibnamefont{and}
  \bibinfo{author}{\bibfnamefont{H.}~\bibnamefont{Asada}},
  \bibinfo{journal}{Universe} \textbf{\bibinfo{volume}{5}},
  \bibinfo{pages}{218} (\bibinfo{year}{2019}), \eprint{1906.02414}.

\end{thebibliography}

\end{document}